\documentclass[sigconf,nonacm]{acmart}

\usepackage{graphicx}
\usepackage{subcaption}
\usepackage{bm}
\usepackage{amsmath}
\usepackage{amsthm}
\usepackage{mathtools}
\usepackage{algorithmicx}
\usepackage{algorithm}
\usepackage[noend]{algpseudocode}
\usepackage[capitalize,noabbrev]{cleveref}
\usepackage{booktabs}
\usepackage{makecell}

\newtheorem{lemma}{Lemma}
\newtheorem{theorem}{Theorem}

\usepackage[most]{tcolorbox}

\newtcolorbox{definition_box}{
    colback=white,
    colframe=black,
    boxrule=0.5pt,
    arc=0pt,
    breakable
}

\newtcolorbox{theorem_box}{
    colback=black!5!white,
    colframe=black!5!white,
    boxrule=0pt,
    arc=0pt,
    breakable
}

\usepackage{enumitem}

\begin{document}

\title[%
Cascaded Learned Bloom Filter for Optimizing Model--Filter Size Balance and Fast Rejection]{%
Cascaded Learned Bloom Filter for%
\texorpdfstring{\\}{ }%
Optimizing Model--Filter Size Balance and Fast Rejection%
}

\author{Atsuki Sato}
\orcid{0009-0001-5366-4842}
\email{a\_sato@hal.t.u-tokyo.ac.jp}
\affiliation{%
    \institution{The University of Tokyo}
    \city{Tokyo}
    \country{Japan}
}

\author{Yusuke Matsui}
\orcid{0000-0003-1529-0154}
\email{matsui@hal.t.u-tokyo.ac.jp}
\affiliation{%
    \institution{The University of Tokyo}
    \city{Tokyo}
    \country{Japan}
}

\renewcommand{\shortauthors}{Sato and Matsui}

\begin{abstract}
Recent studies have demonstrated that learned Bloom filters (LBFs), which combine machine learning with the classical Bloom filter, can achieve superior memory efficiency. 
However, two challenges remain:
(1) jointly optimizing the sizes of the machine learning model and Bloom filters, and
(2) systematically minimizing reject time.
We propose the Cascaded Learned Bloom Filter (CLBF), a unified architecture that generalizes existing LBF designs, including Sandwiched LBF and Partitioned LBF.
Within this framework, we develop a dynamic programming-based optimizer that explores a discretized parameter space and identifies near-optimal configurations that balance model and filter sizes while achieving fast rejection.
Experiments on real-world datasets show that CLBF reduces memory usage by up to 17\% and decreases reject time by up to 65$\times$ compared to Partitioned LBF, the state-of-the-art LBF in terms of memory efficiency under a fixed machine learning model.
\end{abstract}




\maketitle

\section{Introduction}
\label{sec:introduction}

Bloom filters~\cite{bloom1970space} are ubiquitous data structures for approximate membership queries.
A Bloom filter encodes a set $\mathcal{S}$ into a bit array to test whether a query $q$ is a \textit{key} (an element in $\mathcal{S}$) or a \textit{non-key} (an element not in $\mathcal{S}$).
Bloom filters may yield false positives but never false negatives.
Due to their high memory efficiency and fast query performance, they are widely employed in memory-constrained and latency-sensitive applications~\cite{broder2004network, chang2008bigtable}.

Recently, a new class of Bloom filters called Learned Bloom Filters (LBFs) has been proposed~\cite{kraska2018case}.
LBFs leverage a machine learning (ML) model to predict whether an input belongs to the set $\mathcal{S}$, achieving superior memory efficiency compared to classical Bloom filters.
Despite numerous attempts to further improve memory efficiency~\cite{mitzenmacher2018model,dai2020adaptive,vaidya2021partitioned}, two key challenges remain unresolved.

\textbf{(1) Balancing ML model size and Bloom filter size.}
The total memory usage of an LBF is the sum of the ML model size and the Bloom filter size.
A smaller model often (but not always) reduces accuracy and requires larger Bloom filters to meet a target false positive rate (FPR), whereas a larger model can shrink the filters but consumes more memory.
However, most existing LBF construction methods treat the trained model as fixed and only optimize Bloom-filter parameters, leaving the model--filter trade-off unoptimized.

\textbf{(2) Systematically minimizing reject time.}
Bloom filters are often used as pre-filters before an expensive exact membership check.
In such systems, the end-to-end latency of non-key queries decreases as the Bloom filter becomes more accurate and as its \textit{reject time} (i.e., the time required to answer ``$q \notin \mathcal{S}$'') decreases.
Thus, the reject time of a Bloom filter is as important as its accuracy and memory consumption.
However, existing LBF construction methods focus on accuracy and memory, and do not systematically optimize reject time.
In LBFs, reject time depends not only on the cost of Bloom-filter probes but also on the inference cost of the learned component and on how early non-keys can be filtered out along the execution path.
Consequently, minimizing reject time necessitates redesigning the LBF architecture to effectively leverage the ML model and optimize the execution path.

\begin{figure}[t]
    \centering
    \includegraphics[width=0.8\columnwidth]{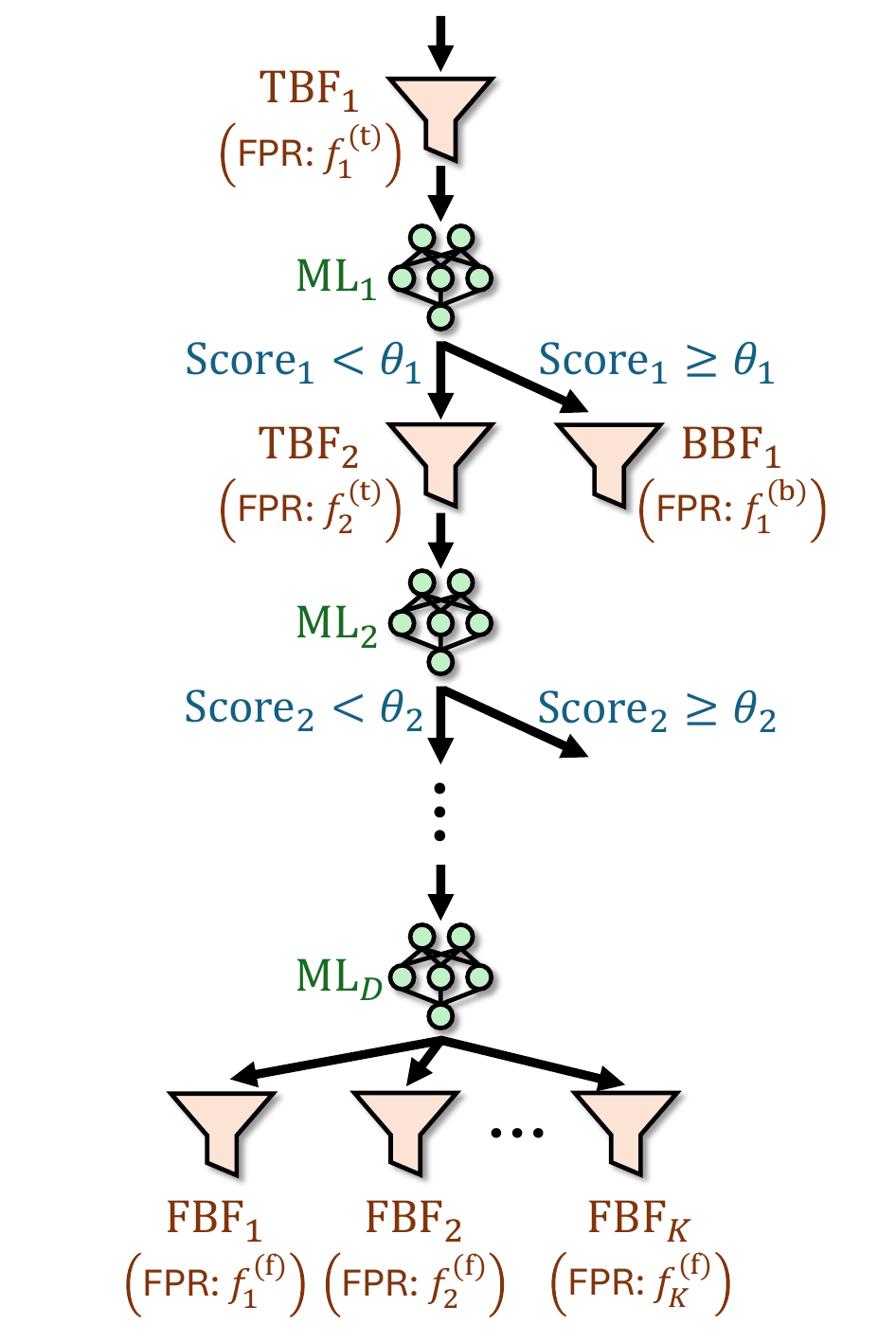}
    \caption{Architecture of the proposed Cascaded Learned Bloom Filter (CLBF).
    CLBF alternates between score-based branching and Bloom filter-based filtering. This design generalizes the architectures of Sandwiched LBF and PLBF.}
    \label{fig:bigmaclbf}
    \Description{Overview of the Cascaded Learned Bloom Filter (CLBF), showing alternating ML models and Bloom filters with score-based branching, generalizing Sandwiched LBF and PLBF.}
\end{figure}

To address these two issues, we propose the \emph{Cascaded Learned Bloom Filter (CLBF)}, which adopts a cascaded architecture with a partitioned structure at the final layer (\cref{fig:bigmaclbf}).
This design generalizes existing LBFs, including naive LBF~\cite{kraska2018case}, Sandwiched LBF~\cite{mitzenmacher2018model}, and Partitioned LBF (PLBF)~\cite{vaidya2021partitioned} (\cref{fig:existinglbf}).
CLBF is built on two key ideas: (1) it first trains a large ML model composed of many weak learners and then prunes it to an appropriate size, thereby allowing the construction to explicitly balance the trade-off between ML model size and Bloom filter size, and (2) it enables faster rejections via branching on tentative model outputs and intermediate Bloom filters.
Our dynamic programming-based optimizer identifies near-optimal configurations (i.e., the ML model size, the thresholds for branching, and each Bloom filter size) over a sufficiently fine-grained discretized parameter space.
Our experiments show that (1) CLBF reduces memory usage by up to 17\% and (2) achieves up to a 65$\times$ reduction in reject time compared to PLBF.


\textbf{Contributions.}
This paper makes the following contributions:
\begin{itemize}[leftmargin=1.5em]
    \item We propose the Cascaded Learned Bloom Filter (CLBF), a unified learned Bloom filter architecture that generalizes existing designs, including naive LBF, Sandwiched LBF, and PLBF.
    \item We formulate CLBF construction as a joint optimization problem over memory usage and expected reject time, where the size of the ML model itself is treated as an explicit optimization variable, rather than being fixed a priori.
    \item We design a dynamic programming-based optimizer that efficiently explores a sufficiently fine-grained discretized parameter space and identifies near-optimal CLBF configurations.
    \item Our extensive experiments demonstrate that, at the same false positive rate, CLBF achieves up to 17\% reduction in memory usage and up to $65\times$ faster rejection compared to PLBF, the state-of-the-art LBF in terms of memory efficiency under a fixed machine learning model.
\end{itemize}

\section{Preliminaries}
\label{sec:prelim}

This section reviews the background of approximate membership queries and learned Bloom filters.
Throughout, $\mathcal{S}$ denotes a set of \emph{keys} and $n \in \mathbb{N}$ denotes the number of keys, i.e., $n = |\mathcal{S}|$.

\subsection{Bloom Filters}
\label{subsec:prelim_bf}

An \emph{approximate membership query} (AMQ) data structure represents a set $\mathcal{S}$ and answers membership queries for an element $q$ by returning either \textsf{Found} (predicting $q\in\mathcal{S}$) or \textsf{NotFound} (predicting $q\notin\mathcal{S}$).
AMQs allow \emph{false positives} but disallow \emph{false negatives}:
\begin{equation}
    \Pr[\textsf{Found}\mid q\notin\mathcal{S}] > 0
    \quad\land\quad
    \Pr[\textsf{NotFound}\mid q\in\mathcal{S}] = 0.
\end{equation}
We refer to $\Pr[\textsf{Found}\mid q\notin\mathcal{S}]$ as the \emph{false positive rate} (FPR).

A classical Bloom filter~\cite{bloom1970space} stores $\mathcal{S}$ in an $m$-bit array using $k$ hash functions.
For a set of size $n$, its FPR is approximately $f \approx (1-e^{-kn/m})^k$, which is minimized when $k \approx \frac{m}{n}\ln 2$.
Under this optimal choice, $f \approx (1/2)^k$, and the bit-array length can be written as $m \approx (\log_2 e)\cdot n\log_2(1/f)$.

\subsection{Learned Bloom Filters}
\label{subsec:prelim_lbf}

\begin{figure}[t]
    \centering
    \includegraphics[width=\columnwidth]{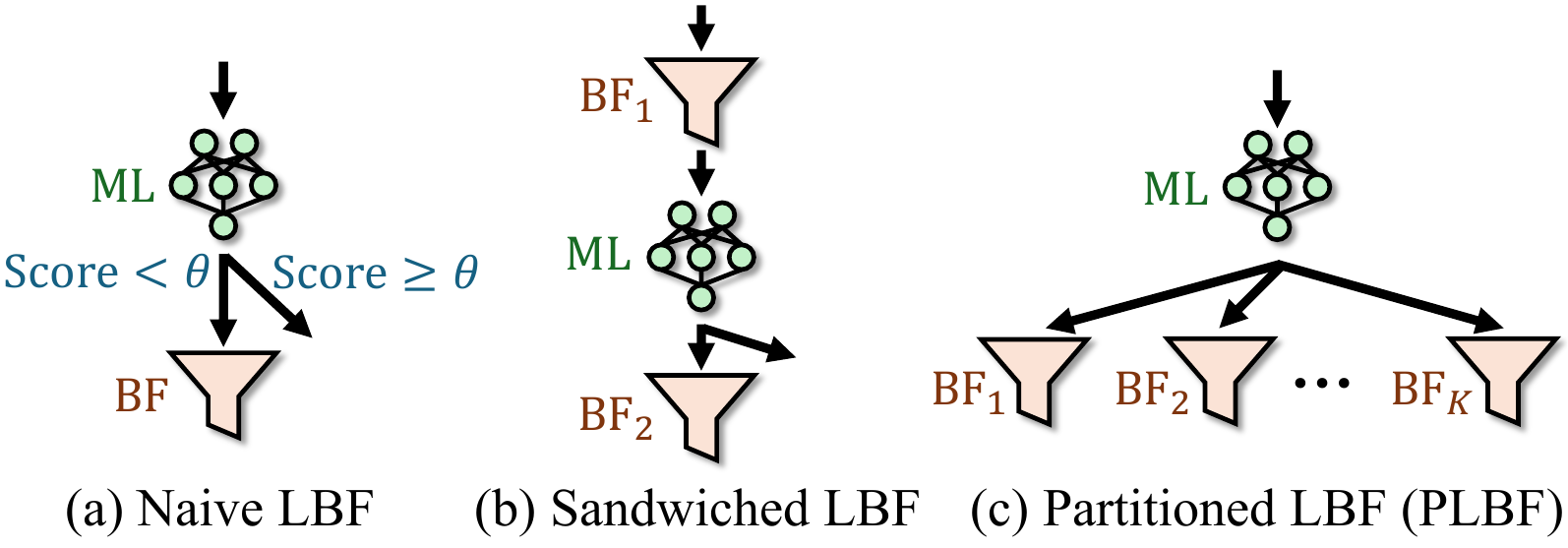}
    \caption{The architecture of existing LBFs: (a) naive LBF~\cite{kraska2018case} has a single backup Bloom filter.
(b) Sandwiched LBF~\cite{mitzenmacher2018model} applies a pre-filter before the model inference.
(c) PLBF~\cite{vaidya2021partitioned} uses multiple backup Bloom filters.}
    \label{fig:existinglbf}
    \Description{Architectures of existing learned Bloom filters: (a) naive LBF with a single backup Bloom filter, (b) Sandwiched LBF with a pre-filter, and (c) PLBF using multiple Bloom filters based on model scores.}
\end{figure}

A learned Bloom filter (LBF)~\cite{kraska2018case} augments an AMQ with a learned model that outputs a score $s(q)\in[0,1]$, representing the model's belief that the filter should answer \textsf{Found} for query $q$, based on the keys and sampled non-keys.
A naive LBF answers a query $q$ by computing $s(q)$, comparing it with a threshold $\theta$, and returning \textsf{Found} if $s(q) \geq \theta$; otherwise, it queries a backup Bloom filter (\cref{fig:existinglbf}a).
To avoid false negatives, the backup filter stores keys rejected by the model, i.e., $\{x\in\mathcal{S}\mid s(x)<\theta\}$.
The threshold $\theta$ and the size of the backup filter are chosen based on the score distribution of keys and sampled non-keys so that the target false positive rate is achieved in expectation.
This design reduces the backup filter size and yields a smaller total memory footprint than a classical Bloom filter.

Sandwiched LBF~\cite{mitzenmacher2018model} adds a Bloom filter \emph{before} the model (\cref{fig:existinglbf}b), and PLBF~\cite{vaidya2021partitioned} uses multiple backup Bloom filters for different score ranges (\cref{fig:existinglbf}c).
These LBFs achieve better memory efficiency than naive LBF by solving a constrained memory minimization problem under a target overall FPR.
In particular, PLBF is the state-of-the-art LBF in terms of memory efficiency when the ML model is fixed.

\subsection{Model--Filter Size Balance of LBFs}
\label{subsec:prelim_balance}

A key design goal of LBFs is to minimize the \emph{total memory usage}, i.e., the sum of the memory usage of the model and that of the Bloom-filter components, while satisfying a target FPR.
These components are coupled: a smaller model often reduces prediction accuracy, which increases the backup filter size and/or the model-induced false positives, requiring larger Bloom filters to meet a target FPR.
Conversely, a larger model can shrink the filters but may dominate memory.
Most prior LBF construction pipelines fix the model architecture/size and optimize only Bloom-filter parameters (e.g., thresholds and filter sizes), leaving this model--filter trade-off unoptimized and potentially far from optimal.

\subsection{Reject Time of LBFs}
\label{subsec:prelim_reject}

In storage engines, AMQ structures are commonly used as prefilters before expensive \emph{exact} membership checks (e.g., on-disk lookups)~\cite{broder2004network, chang2008bigtable}.
These systems motivate an explicit focus on \emph{reject time}: the latency to conclude that a query $q\notin\mathcal{S}$.
Let $T_{\mathrm{strict}}$ be the latency of an exact membership check without any prefilter.
Let $f$ be the FPR of the prefiltering structure, and let $t_{\mathrm{acc}}$ and $t_{\mathrm{rej}}$ be its \emph{accept time} and \emph{reject time}, respectively.
For a non-key query, the expected end-to-end latency of an exact membership check is
\begin{equation}
    \label{equ: approx time for non-key query}
    (1-f) t_{\mathrm{rej}} + f (t_{\mathrm{acc}} + T_{\mathrm{strict}}) \approx t_{\mathrm{rej}} + fT_{\mathrm{strict}},
\end{equation}
since typically $t_{\mathrm{acc}}\ll T_{\mathrm{strict}}$ and $f\ll 1$.
This latency is typically much smaller than $T_{\mathrm{strict}}$, which is why prefiltering is employed.
Moreover, \cref{equ: approx time for non-key query} shows that minimizing end-to-end latency requires not only reducing $f$, but also aggressively reducing $t_{\mathrm{rej}}$.

For LBFs, however, controlling $t_{\mathrm{rej}}$ is fundamentally more challenging than for classical AMQs.
Reject time is no longer determined solely by Bloom-filter probes;
it also depends on the inference cost of the learned model and on how early non-keys can be filtered out along the execution path (e.g., by a prefilter or an ML model).
Consequently, two LBF designs with identical memory usage and FPR can exhibit vastly different reject times, depending on how the learned component is organized and how queries flow through it.

An LBF construction that achieves both low reject time and high memory efficiency must open the ML black box just enough to optimize the execution path:
it should reject easy non-keys early using inexpensive components, while invoking more costly learned computation only when it improves memory efficiency or FPR.
Simply replacing the learned model with a very fast but inaccurate one reduces reject time but destroys the memory--accuracy advantage of LBFs.
The key design challenge is therefore to make the latency of the learned component controllable while keeping the framework general enough to exploit expressive models when they are beneficial.

Despite this challenge, most prior LBF studies primarily optimize accuracy or memory efficiency and do not explicitly consider reject time as an optimization objective~\cite{kraska2018case,mitzenmacher2018model,vaidya2021partitioned}.
A few recent studies discuss practical guidelines for choosing AMQ variants or operating points under latency considerations, but they provide neither a general, automatic procedure for minimizing reject time under memory constraints, nor a unified construction framework for jointly optimizing memory efficiency and reject time~\cite{fumagalli2022choice,malchiodi2024role}.

\section{Overview of CLBF}
\label{sec:overview_of_clbf}

In this section, we first describe the core design concepts behind our CLBF, and then discuss the overall architecture and workflow.
The main notations are summarized in \cref{tab:notations}.

\subsection{Core Design Concepts}
\label{sec:core_design_concepts}

\begin{table}[t]
    \centering
    \caption{Major Notations.}
    \label{tab:notations}
    \setlength{\tabcolsep}{3pt}
    \small
    \begin{tabular}{c|l} \toprule
    \multicolumn{2}{l}{\textbf{\cref{sec:overview_of_clbf}: Overview of CLBF}} \\ \midrule
    $D \in\mathbb{N}$ & Number of ML models to use \\
    $K \in\mathbb{N}$ & Number of Final Bloom filters \\
    $\mathrm{TBF}_{d}$ & $d$-th Trunk Bloom Filter ($d\in\{1,\dots,D\}$)\\
    $\mathrm{BBF}_{d}$ & $d$-th Branch Bloom Filter ($d\in\{1,\dots,D-1\}$)\\
    $\mathrm{FBF}_{k}$ & $k$-th Final Bloom Filter ($k\in\{1,\dots,K\}$)\\
    $\mathrm{ML}_{d}$ & $d$-th ML model ($d\in\{1,\dots,D\}$)\\
    $\bm{\theta}\in\mathbb{R}^{D-1}$ & $\theta_{d}$: threshold for $\mathrm{ML}_{d}$'s score ($d\in\{1,\dots,D-1\}$)\\
    $\bm{f}^{(\mathrm{t})} \in(0,1]^D$& $f^{(\mathrm{t})}_d$: FPR of $\mathrm{TBF}_{d}$ ($d\in\{1,\dots,D\}$)\\
    $\bm{f}^{(\mathrm{b})} \in(0,1]^{D-1}$ & $f^{(\mathrm{b})}_d$: FPR of $\mathrm{BBF}_{d}$ ($d\in\{1,\dots,D-1\}$)\\
    $\bm{f}^{(\mathrm{f})} \in(0,1]^K$ & $f^{(\mathrm{f})}_k$: FPR of $\mathrm{FBF}_{k}$ ($k\in\{1,\dots,K\}$)\\
    $\mathcal{F}$ & All FPRs, i.e., $\{\bm{f}^{(\mathrm{t})},\bm{f}^{(\mathrm{b})},\bm{f}^{(\mathrm{f})}\}$ \\ \midrule

    \addlinespace[2pt]
    \multicolumn{2}{l}{\textbf{\cref{sec:problem_for_clbf_construction}: CLBF Configuration Problem}} \\ \midrule
    $\mathcal{S}$ & Set of keys \\
    $\mathcal{Q}$ & Set of non-keys \\
    $n\in\mathbb{N}$ & Number of keys, i.e., $|\mathcal{S}|$ \\
    $\bar{D} \in\mathbb{N}$ & Number of given ML models ($\bar{D} \geq D$) \\
    $F \in(0,1)$ & Target FPR \\
    $\lambda \in[0,1]$ & Trade-off parameter for memory and reject time \\
    $\mathcal{L}(D,\mathcal{F},\bm{\theta})$ & Objective function to be minimized (\cref{eq:objective}) \\
    $M(D,\mathcal{F},\bm{\theta})$ & Memory usage of CLBF (\cref{eq:memory_size}) \\
    $R(D,\mathcal{F},\bm{\theta})$ & Reject time of CLBF (\cref{eq:reject_time}) \\
    $M_\mathrm{BF}\in\mathbb{R}_{>0}$ & Memory usage of a Bloom filter ($n$ keys, FPR $F$) \\
    $R_\mathrm{BF}\in\mathbb{R}_{>0}$ & Reject time of a Bloom filter ($n$ keys, FPR $F$) \\
    $g^{(\mathrm{t})}_d, g^{(\mathrm{b})}_d, g^{(\mathrm{f})}_{D, k}$ & Proportion of keys reaching $\mathrm{TBF}_d, \mathrm{BBF}_d, \mathrm{FBF}_k$ \\
    $h^{(\mathrm{t})}_d, h^{(\mathrm{b})}_d, h^{(\mathrm{f})}_{D, k}$ & Proportion of non-keys reaching $\mathrm{TBF}_d, \mathrm{BBF}_d, \mathrm{FBF}_k$ \\
    $\mathrm{Size}(\cdot)$ & Memory size of the input \\
    $\mathrm{Time}(\cdot)$ & Inference time of the input \\ \midrule

    \addlinespace[2pt]
    \multicolumn{2}{l}{\textbf{\cref{sec:clbf_configuration_optimization}: CLBF Configuration Optimization}} \\ \midrule
    $\mathcal{L}_d(D,\mathcal{F})$ & Objective function under $\mathrm{TBF}_d$ (\cref{sec:appendix:L_d_def}) \\
    $\mathrm{dp}(d, T)$ & $\min \mathcal{L}_d \text{~s.t.~} \prod_{j=1}^{d-1} f^{(\mathrm{t})}_j = T$ (\cref{eq:dp}) \\
    $\hat{\mathrm{dp}}(d, T)$ & $\min \mathcal{L}_d \text{~s.t.~} \prod_{j=1}^{d-1} f^{(\mathrm{t})}_j = T ~ \land ~ D = d$ \\
    $\check{\mathrm{dp}}(d, T)$ & $\min \mathcal{L}_d \text{~s.t.~} \prod_{j=1}^{d-1} f^{(\mathrm{t})}_j = T ~ \land ~ D > d$ \\ 
    $P \in\mathbb{N}$ & Discretization parameter ($f^{(t)}_d, T \in \{0.5^i\}_{i=0}^{P-1}$) \\ \bottomrule
    \end{tabular}
\end{table}

\begin{figure*}[t]
    \centering
    \begin{minipage}[t]{0.48\linewidth}
        \vspace{0pt}
        \begin{algorithm}[H]
        \caption{Query Processing in CLBF}
        \label{alg:query_processing}
        \begin{algorithmic}[1]
        \State \textbf{Input:} Query $q$, \;\; \textbf{Output:} $\texttt{NotFound}$ or $\texttt{Found}$
        \vspace{0.2em}
        \For{$d = 1, 2, \dots, D - 1$}
            \If{$\mathrm{TBF}_d.\mathrm{Lookup}(q) = \texttt{NotFound}$} \textbf{return} $\texttt{NotFound}$ \EndIf
            \vspace{-0.1em}
            \State $s \gets \mathrm{ML}_d(q)$
            \If{$s \geq \theta_d$}
                \State \textbf{return} $\mathrm{BBF}_d.\mathrm{Lookup}(q)$
            \EndIf
        \EndFor
        \If{$\mathrm{TBF}_D.\mathrm{Lookup}(q) = \texttt{NotFound}$} \textbf{return} $\texttt{NotFound}$ \EndIf
        \vspace{-0.1em}
        \State $s \gets \mathrm{ML}_D(q)$
        \State $k \gets \mathrm{GetFBFIndex}(s)$  \Comment{gets the index of $\mathrm{FBF}$ for score $s$}
        \State \textbf{return} $\mathrm{FBF}_k.\mathrm{Lookup}(q)$
        \end{algorithmic}
        \end{algorithm}
    \end{minipage}
    \hfill
    \begin{minipage}[t]{0.48\linewidth}
        \vspace{0pt}
        \begin{algorithm}[H]
        \caption{Key Insertion into CLBF}
        \label{alg:key_insertion}
        \begin{algorithmic}[1]
        \State \textbf{Input:} Key $q$
        \vspace{0.4em}
        \For{$d = 1, 2, \dots, D - 1$}
            \State $\mathrm{TBF}_d.\mathrm{Insert}(q)$
            \State $s \gets \mathrm{ML}_d(q)$
            \If{$s \geq \theta_d$}
                \State $\mathrm{BBF}_d.\mathrm{Insert}(q); ~ \textbf{return}$
            \EndIf
        \EndFor
        \State $\mathrm{TBF}_D.\mathrm{Insert}(q)$
        \State $s \gets \mathrm{ML}_D(q)$
        \State $k \gets \mathrm{GetFBFIndex}(s)$ \Comment{gets the index of $\mathrm{FBF}$ for score $s$}
        \State $\mathrm{FBF}_k.\mathrm{Insert}(q)$
        \end{algorithmic}
        \end{algorithm}
    \end{minipage}
    \Description{Two pseudocode listings for CLBF: left shows query processing with early exits via trunk and branch Bloom filters; right shows key insertion along the same path to avoid false negatives.}
\end{figure*}

CLBF optimizes both the Bloom-filter component and the learned component to jointly minimize \emph{memory usage} and \emph{reject time} under a target FPR. To this end, CLBF follows the design principles below.

\textbf{Expose model size as an optimization variable via an expandable ensemble.}
Rather than treating the learned component as a single fixed black box, CLBF represents it as an ordered sequence of modular base learners, often referred to as weak learners in boosting.
These weak learners may be decision trees, small neural networks, or other inexpensive prediction modules.
Using more weak learners increases the model's memory footprint, but typically improves the predictive accuracy of the ensemble.
At construction time, CLBF selects an appropriate number of weak learners to include.
This makes the model footprint a tunable parameter that can be traded against Bloom-filter memory, enabling a near-optimal balance between model capacity and filter allocation.

\textbf{Cascade inexpensive tests before expensive ones.}
The reject time for non-key queries largely depends on how quickly they can be eliminated before reaching expensive computations.
CLBF therefore interleaves inexpensive Bloom-filter probes with learned predictions and supports \emph{early exits} based on tentative model scores (\cref{fig:bigmaclbf}).
After an intermediate model produces a score, a query proceeds to a deeper stage if the current prediction is not sufficiently confident.
When the prediction is sufficiently confident, however, the query can instead be routed to a cheaper verification path.
Intermediate Bloom filters also serve to filter out non-key queries early, thereby reducing reject time.

\textbf{Exploit score heterogeneity with a partitioned final backup.}
Even among queries that survive to the final stage, the difficulty of distinguishing keys from non-keys may vary substantially, as reflected by their learned scores.
CLBF therefore employs multiple final Bloom filters for different score regions (as in PLBF~\cite{vaidya2021partitioned}), enabling finer-grained memory allocation and more effective use of model confidence than a single backup filter.

\textbf{Unify and generalize prior LBF architectures in one framework.}
By combining a cascaded structure of models and Bloom filters with a partitioned final backup, CLBF forms a generalized LBF architecture.
Through appropriate parameter choices (e.g., the FPRs of the Bloom filters, the thresholds of intermediate models, and the number of final Bloom filters), CLBF recovers common LBF variants as special cases while enlarging the design space (\cref{fig:bigmaclbf,fig:existinglbf}).
This unified formulation allows a single optimizer to explore architectures that differ in memory usage and reject time.

\subsection{Overall Architecture and Workflow}
\label{sec:overall_architecture_and_workflow}

CLBF employs a cascade structure consisting of multiple ML models and multiple Bloom filters arranged alternately (\cref{fig:bigmaclbf}).
Here, we first describe how queries are processed in CLBF (\cref{alg:query_processing}) and how keys are inserted into the structure (\cref{alg:key_insertion}), while introducing the notation used throughout.

\textbf{Query Processing in CLBF (\cref{alg:query_processing}).}
At each intermediate depth $d \in \{1, \dots, D-1\}$, the query first probes a \emph{Trunk Bloom filter} ($\mathrm{TBF}_d$).
A $\mathrm{TBF}$ is placed \emph{before} the corresponding ML model to cheaply reject non-key queries and avoid unnecessary inference.
If $\mathrm{TBF}_d$ returns \textsf{NotFound}, CLBF terminates immediately with \textsf{NotFound}.
Otherwise, CLBF passes the query to $\mathrm{ML}_d$, which outputs a score $s_d$.
If $s_d \geq \theta_d$, the intermediate model is sufficiently confident, and CLBF takes the \emph{branch} path: a \emph{Branch Bloom filter} ($\mathrm{BBF}_d$) makes the final decision without proceeding to deeper stages.
This early exit avoids the cost of deeper models and filters for queries that can already be resolved at depth $d$.
If $s_d < \theta_d$, the query is not yet confidently classified and proceeds to the next depth.
CLBF repeats this pattern across depths.
At the final model ($d=D$), CLBF follows the PLBF-style partitioned backup.
Based on the range in which the final score falls, one of the $K$ \emph{Final Bloom filters} ($\mathrm{FBF}_1, \mathrm{FBF}_2, \dots, \mathrm{FBF}_K$) is selected to make the final determination.
Each $\mathrm{FBF}$ serves as a score-conditioned backup filter, enabling finer-grained memory allocation across score regions than a single final filter.

Note that the three types of Bloom filters (i.e., $\mathrm{TBF}$, $\mathrm{BBF}$, and $\mathrm{FBF}$) in CLBF differ only in their roles;
their underlying implementations need not be different.
In our experiments, we instantiate all filters as standard Bloom filters, but they can be replaced by any classical AMQ data structure, such as cuckoo filters or quotient filters.
Similarly, CLBF can employ arbitrary ML models.
In our main experiments (\cref{sec:experiments}), each ML model in CLBF corresponds to a weak learner trained using a boosting algorithm.

\textbf{Key Insertion into CLBF (\cref{alg:key_insertion}).}
When inserting a key into CLBF, we follow the same path as in query processing and insert the key into all Bloom filters along that path.
This guarantees that each Bloom filter along the path confirms the key's presence, thereby ensuring that CLBF produces no false negatives.

\textbf{Generalization of Existing LBF Architectures.}
The CLBF architecture generalizes existing LBF architectures (\cref{fig:existinglbf,fig:bigmaclbf}):
with a single TBF and a single FBF it reduces to a Sandwiched LBF, and without TBFs/BBFs but with multiple FBFs it reduces to a PLBF.
Under this unified framework, we optimize both the number of ML models, the Bloom-filter FPRs, and thresholds, achieving better trade-offs among memory usage, reject time, and accuracy compared to existing LBF architectures.

\textbf{Notations for CLBF Configuration.}
$D$ is the number of ML models in CLBF (optimized via dynamic programming).
Let $f^{(\mathrm{t})}_d$ ($d=1,2,\dots,D$), $f^{(\mathrm{b})}_d$ ($d=1,2,\dots,D-1$), and $f^{(\mathrm{f})}_k$ ($k=1,2,\dots,K$) denote the FPRs of $\mathrm{TBF}_d$, $\mathrm{BBF}_d$, and $\mathrm{FBF}_k$, respectively.
For convenience, we write these in vector form, e.g., $\bm{f}^{(\mathrm{t})}=[f^{(\mathrm{t})}_1,\dots,f^{(\mathrm{t})}_D]$, and write $\mathcal{F}=\{\bm{f}^{(\mathrm{t})},\bm{f}^{(\mathrm{b})},\bm{f}^{(\mathrm{f})}\}$.

\section{CLBF Configuration Problem}
\label{sec:problem_for_clbf_construction}


\begin{algorithm}[t]
    \caption{Construction of CLBF}
    \label{alg:clbf_construction}
    \begin{algorithmic}[1]
    \State \textbf{Input:} $\mathcal{S}, \mathcal{Q}, F, \lambda$
    \vspace{0.5em}
    \State $\mathrm{ML}_1, \mathrm{ML}_2, \dots, \mathrm{ML}_{\bar{D}} \gets \mathrm{TrainML}(\mathcal{S}, \mathcal{Q})$
    \State $M_\mathrm{BF}, R_\mathrm{BF} \gets \mathrm{BenchmarkClassicBF}(\mathcal{S},\mathcal{Q},F)$\label{line:benchmarkclassicbf}
    \State $D^{\ast},\mathcal{F}^{\ast},\bm{\theta}^{\ast}, L^{\ast} \gets \texttt{None}, \texttt{None}, \texttt{None}, \infty$
    \While{\textbf{true}}
        \State $\bm{\theta} \gets \mathrm{GetNextTheta}(\mathcal{S}, \mathcal{Q}, \mathrm{ML}_{1:\bar{D}})$ \Comment{Sec. \ref{sec:optimizing_intermediate_thresholds}}\label{line:getnexttheta}
        \If{$\bm{\theta} \mathrm{~is~} \texttt{None}$} \textbf{break} \EndIf\label{line:ifthetaisnonebreak}
        \State $D, \mathcal{F} \gets \mathrm{OptDF}(\mathcal{S}, \mathcal{Q}, F, \lambda, \bm{\theta}, M_\mathrm{BF}, R_\mathrm{BF}, \mathrm{ML}_{1:\bar{D}})$ \Comment{Sec. \ref{sec:dynamic_programming_for_optimizing_d_and_f}}\label{line:optdf}
        \If{$\mathcal{L}(D,\mathcal{F},\bm{\theta}) < L^{\ast}$} \Comment{Eq. \ref{eq:objective}}
            \State $D^{\ast},\mathcal{F}^{\ast},\bm{\theta}^{\ast}, L^{\ast} \gets D,\mathcal{F},\bm{\theta}, \mathcal{L}(D,\mathcal{F},\bm{\theta})$
        \EndIf
    \EndWhile
    \State $\mathrm{CLBF} \gets \mathrm{InitializeCLBF}(\mathcal{F}^{\ast},\bm{\theta}^{\ast},\mathcal{S},\mathrm{ML}_{1:D^{\ast}})$
    \For{$s \in \mathcal{S}$}
        \State $\mathrm{CLBF}.\mathrm{Insert}(s)$ \Comment{Alg. \ref{alg:key_insertion}}
    \EndFor
    \end{algorithmic}
\end{algorithm}

The goal of CLBF construction is to minimize the weighted sum of memory usage and the expected reject time, subject to an accuracy constraint.
We first describe the inputs to CLBF construction, then introduce the parameters to be optimized, and finally present the objective function.
The entire construction pipeline is summarized in \cref{alg:clbf_construction}.
This section focuses on the design principles of CLBF construction, and defers the detailed optimization procedure (i.e., the process inside the \textbf{while} loop in \cref{alg:clbf_construction}) to \cref{sec:clbf_configuration_optimization}.

\textbf{Inputs for CLBF Construction.}
To construct the CLBF, the user must provide the following four components:
\begin{itemize}[leftmargin=1.5em]
    \item \textbf{Key and non-key sets.}
        As in prior LBFs~\cite{kraska2018case,mitzenmacher2018model,vaidya2021partitioned}, the construction requires not only a set of keys $\mathcal{S}$ but also a set of non-keys $\mathcal{Q}$ such that $\mathcal{S} \cap \mathcal{Q} = \varnothing$.
        The non-key set $\mathcal{Q}$ is used both to train the ML models and to simulate the behavior of non-key queries during CLBF evaluation.
    \item \textbf{Target FPR.} 
        The target FPR is denoted by $F \in (0, 1)$.
        The CLBF is optimized to minimize memory usage and reject time, subject to the constraint that the expected FPR does not exceed $F$. 
    \item \textbf{Trade-off parameter.} 
        The parameter $\lambda \in [0, 1]$ controls the trade-off between memory efficiency and reject time.
        As $\lambda$ increases, the optimization algorithm focuses more on reducing memory usage; specifically, when $\lambda = 1$, only memory usage is minimized, while when $\lambda = 0$, only reject time is minimized.
\end{itemize}

The sets $\mathcal{S}$ and $\mathcal{Q}$ (or their subsets) are used as \textit{training data} to train multiple ML models for binary classification between keys and non-keys.
The number of trained models is denoted by $\bar{D}$.
Among these models $\mathrm{ML}_1, \dots, \mathrm{ML}_{\bar{D}}$, only the first $D ~ (\leq \bar{D})$ models, i.e., $\mathrm{ML}_1, \mathrm{ML}_2, \dots, \mathrm{ML}_D$, are used in the CLBF.
The value of $D$ is optimized during CLBF construction, as described in \cref{sec:clbf_configuration_optimization}.

In addition, the sets $\mathcal{S}$ and $\mathcal{Q}$ (or their subsets) are used as \textit{calibration data} to trace the flow of keys and non-keys through the CLBF structure, measuring the proportion that reaches each ML model and Bloom filter.
In addition, the calibration data is used to evaluate the inference time of each ML model and estimate the overall reject time of the CLBF.






\textbf{Parameters to be Optimized.}
To construct the CLBF, we optimize the number of ML models used ($D$), the FPRs of the Bloom filters ($\mathcal{F}=\{\bm{f}^{(\mathrm{t})}, \bm{f}^{(\mathrm{b})}, \bm{f}^{(\mathrm{f})}\}$), and the intermediate thresholds ($\bm{\theta}=[\theta_1,\theta_2,\dots,\theta_{D-1}]$).
The thresholds $\bm{\theta}$ are explored in the outer loop via $\mathrm{GetNextTheta}$ (Line~\ref{line:getnexttheta} of \cref{alg:clbf_construction}), as detailed in \cref{sec:optimizing_intermediate_thresholds}.
For each candidate $\bm{\theta}$, the optimal $D$ and $\mathcal{F}$ are then computed in the inner optimization (Line~\ref{line:optdf}) using dynamic programming, as described in \cref{sec:dynamic_programming_for_optimizing_d_and_f}.
Although the thresholds for the final model $\mathrm{ML}_D$ are also a target of optimization, their optimal values can be obtained using the method proposed in PLBF~\cite{vaidya2021partitioned}.
This method selects the thresholds that maximize the KL divergence between the score distributions of keys and non-keys. Therefore, we focus on optimizing $D$, $\mathcal{F}$, and $\bm{\theta}$.

Note that once $D$, $\mathcal{F}$, and $\bm{\theta}$ are fixed, the FPR and the number of inserted keys for each Bloom filter (i.e., the $\mathrm{TBF}$s, $\mathrm{BBF}$s, and $\mathrm{FBF}$s) are determined, which in turn specify its configuration, such as the bit-array length and the number of hash functions.
A Bloom filter's configuration is uniquely determined by its target FPR and the number of inserted keys.

\textbf{Objective Function and Constraint.}
The following objective function $\mathcal{L}(D,\mathcal{F},\bm{\theta})$ is minimized under the constraint that the \textit{expected FPR} does not exceed $F$:
\begin{equation}
    \label{eq:objective}
    \mathcal{L}(D,\mathcal{F},\bm{\theta}) \coloneq \lambda \cdot \frac{M(D,\mathcal{F},\bm{\theta})}{M_\mathrm{BF}} + (1-\lambda) \cdot \frac{R(D,\mathcal{F},\bm{\theta})}{R_\mathrm{BF}},
\end{equation}
where $M(D,\mathcal{F},\bm{\theta})$ represents the \textit{memory usage} of the CLBF, and $R(D,\mathcal{F},\bm{\theta})$ denotes the \textit{expected reject time}.
The quantities $M_\mathrm{BF}$ and $R_\mathrm{BF}$ are normalization constants: $M_\mathrm{BF}$ is the memory usage of a classical Bloom filter configured with the same number of keys $n$ and target FPR $F$, and $R_\mathrm{BF}$ is its reject time.
We measure $M_\mathrm{BF}$ and $R_\mathrm{BF}$ using the calibration data (Line~\ref{line:benchmarkclassicbf} of \cref{alg:clbf_construction}).

\begin{figure}[t]
    \centering
    \includegraphics[width=\columnwidth]{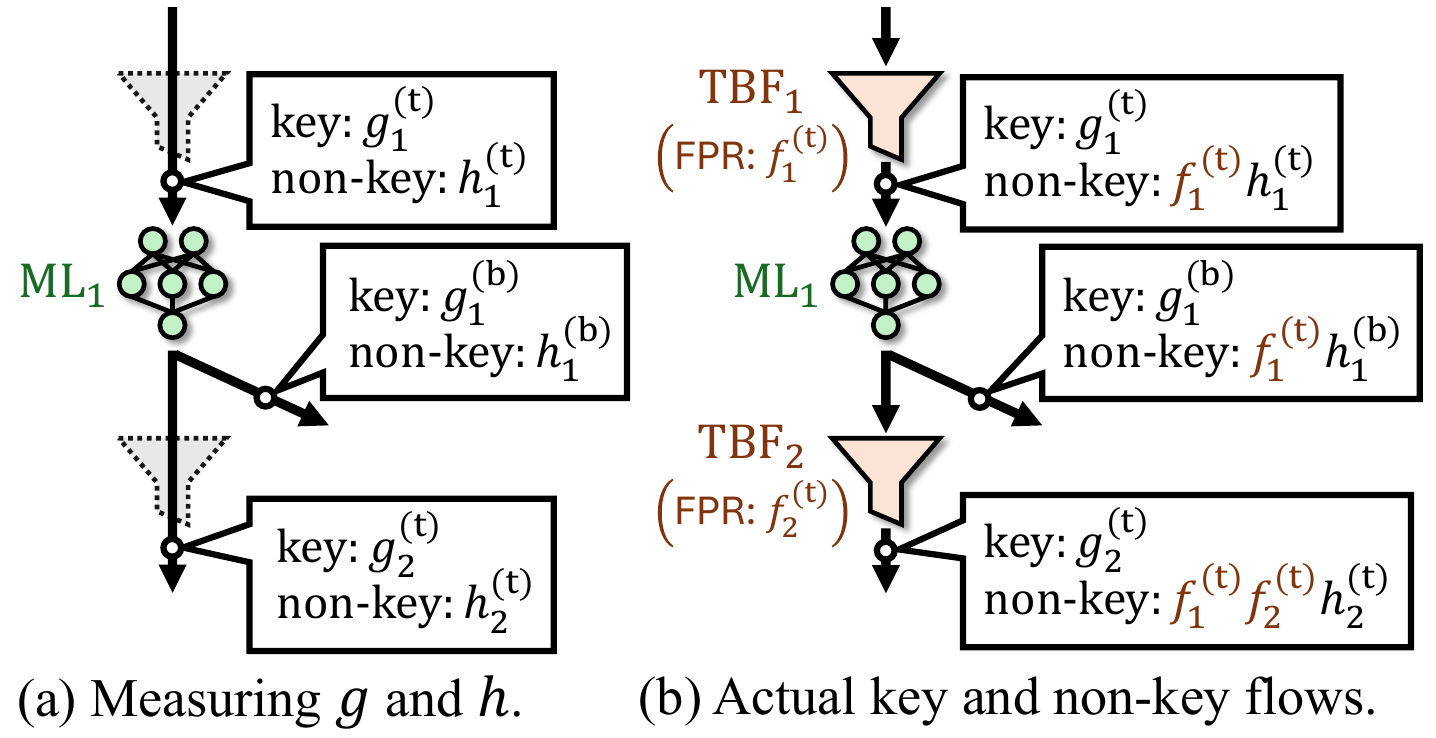}
    \caption{(a) When measuring $g$ and $h$, we omit TBF filtering. (b) With TBFs, the key fractions remain unchanged because TBFs have no false negatives, whereas the non-key fraction reaching each stage is multiplied by the product of the FPRs of all upstream TBFs, since only non-keys that pass every upstream TBF as false positives can reach that stage.}
    \label{fig:nonkey_ratio}
\end{figure}

\textbf{How to Evaluate Each Term.}
Once the thresholds $\bm{\theta}$ are fixed, we measure the flow of keys and non-keys through the CLBF structure using the calibration data, \textbf{without} applying the filtering by the $\mathrm{TBF}$s, i.e., considering only score-based branching (\cref{fig:nonkey_ratio}(a)).
We define $g^{(\mathrm{t})}_{d}, g^{(\mathrm{b})}_{d}, g^{(\mathrm{f})}_{D,k} \in [0,1]$ (resp.\ $h^{(\mathrm{t})}_{d}, h^{(\mathrm{b})}_{d}, h^{(\mathrm{f})}_{D,k} \in [0,1]$) as the proportions of keys (resp.\ non-keys) in the calibration data passed to $\mathrm{TBF}_d$, $\mathrm{BBF}_d$, and $\mathrm{FBF}_k$, respectively.
Note that, since the calibration data and $\bm{\theta}$ are fixed, these proportions are deterministic quantities; they depend on $\bm{\theta}$, but we omit this dependence from the notation for brevity.
In contrast, the false positives of the Bloom filters are random events determined by their hash functions.
We assume that, for every non-key query, the false-positive events of distinct Bloom filters are mutually independent.
This assumption is natural because each filter is built with its own independently chosen hash functions; hence, whether a query passes one filter as a false positive carries no information about whether it passes another.

Under this model, the following lemma formalizes how the filtering by the $\mathrm{TBF}$s changes the query flow, as illustrated in \cref{fig:nonkey_ratio}(b).
\begin{theorem_box}
\begin{lemma}[Query Flow under TBF Filtering]
    \label{lem:query_flow}
    Assume that for every non-key query, the lookup of $\mathrm{TBF}_d$ returns \textsf{Found} with probability $f^{(\mathrm{t})}_d$ for each $d \in \{1,\dots,D\}$, and for every non-key query, these events are mutually independent across $\mathrm{TBF}_1, \dots, \mathrm{TBF}_D$.
    Then, for each depth $d$ and each $k \in \{1,\dots,K\}$, when the filtering by the $\mathrm{TBF}$s is applied,
    (i) the proportions of keys entering $\mathrm{TBF}_d$, $\mathrm{BBF}_d$, and $\mathrm{FBF}_k$ are exactly $g^{(\mathrm{t})}_d$, $g^{(\mathrm{b})}_d$, and $g^{(\mathrm{f})}_{D,k}$, respectively, and
    (ii) the expected proportions of non-keys entering $\mathrm{ML}_d$, $\mathrm{BBF}_d$, and $\mathrm{FBF}_k$ are $h^{(\mathrm{t})}_d \prod_{j=1}^{d} f^{(\mathrm{t})}_j$, $h^{(\mathrm{b})}_d \prod_{j=1}^{d} f^{(\mathrm{t})}_j$, and $h^{(\mathrm{f})}_{D,k} \prod_{j=1}^{D} f^{(\mathrm{t})}_j$, respectively.
\end{lemma}
\end{theorem_box}
\begin{proof}
(i) $\mathrm{TBF}$s never remove a key from its path; each key thus follows the same deterministic path as under the score-based branching alone, and the key proportions coincide with those measured in \cref{fig:nonkey_ratio}(a).
(ii) We show the claim for $\mathrm{ML}_d$; the other cases are analogous.
Whether the score-based branching alone routes a given non-key query down to depth $d$ is deterministic, and the fraction of non-keys for which this holds is $h^{(\mathrm{t})}_d$ by definition.
Under TBF filtering, such a query enters $\mathrm{ML}_d$ if and only if it additionally passes $\mathrm{TBF}_1, \dots, \mathrm{TBF}_d$ as false positives, which occurs with probability $\prod_{j=1}^{d} f^{(\mathrm{t})}_j$ by assumption; any other non-key query never enters $\mathrm{ML}_d$.
Hence, each non-key enters $\mathrm{ML}_d$ with probability $\prod_{j=1}^{d} f^{(\mathrm{t})}_j$ if it is routed to depth $d$ by the score-based branching, and with probability $0$ otherwise.
By linearity of expectation, the expected proportion of non-keys entering $\mathrm{ML}_d$ is therefore $h^{(\mathrm{t})}_d \prod_{j=1}^{d} f^{(\mathrm{t})}_j$.
\end{proof}

We can compute $M(D,\mathcal{F},\bm{\theta})$ by summing the memory costs of all components:
\begin{align}
\label{eq:memory_size}
    M(D,\mathcal{F},\bm{\theta}) &\coloneq \sum_{i=1}^{D} \mathrm{Size}(\mathrm{ML}_i) + \sum_{i=1}^{D} \mathrm{Size}(\mathrm{TBF}_i) + \nonumber \\
    &\qquad + \sum_{i=1}^{D-1} \mathrm{Size}(\mathrm{BBF}_i) + \sum_{k=1}^{K} \mathrm{Size}(\mathrm{FBF}_k).
\end{align}
Since a Bloom filter storing $n$ keys with FPR $f$ requires $c n \log_2(1/f)$ bits, where $c=\log_2(e)$, the memory costs of the Bloom-filter components can be written more explicitly.
For example, by \cref{lem:query_flow}~(i), $\mathrm{TBF}_i$ stores $n g^{(\mathrm{t})}_i$ keys and has FPR $f^{(\mathrm{t})}_i$. Its memory cost is therefore
\begin{equation}
\label{eq:bloom_filter_size_1}
    \mathrm{Size}(\mathrm{TBF}_i) = c n g^{(\mathrm{t})}_i \log_2 (1/f^{(\mathrm{t})}_i).
\end{equation}
Similarly, the Bloom-filter terms in \cref{eq:memory_size} are given by
\begin{align}
\label{eq:bloom_filter_size_2}
    \mathrm{Size}(\mathrm{BBF}_i) &= c n g^{(\mathrm{b})}_i \log_2 (1/f^{(\mathrm{b})}_i), \\
\label{eq:bloom_filter_size_3}
    \mathrm{Size}(\mathrm{FBF}_k) &= c n g^{(\mathrm{f})}_{D,k} \log_2 (1/f^{(\mathrm{f})}_k).
\end{align}

To evaluate $R(D,\mathcal{F},\bm{\theta})$, we make the following two assumptions (both of which are often supported empirically): (i) for each ML model, the time required to evaluate a single non-key query can be treated as constant across queries; and (ii) the reject time can be approximated by the total time spent on inference by the ML models. Under these assumptions, $R(D,\mathcal{F},\bm{\theta})$ can be computed as follows:
\begin{align}
\label{eq:reject_time}
    R(D,\mathcal{F},\bm{\theta}) \coloneq \sum_{i=1}^{D} \left(\mathrm{Time}(\mathrm{ML}_i) \cdot h^{(\mathrm{t})}_i \prod_{j=1}^{i} f^{(\mathrm{t})}_j\right).
\end{align}
By \cref{lem:query_flow}~(ii), the factor $h^{(\mathrm{t})}_i \prod_{j=1}^{i} f^{(\mathrm{t})}_j$ is the expected proportion of non-key queries processed by $\mathrm{ML}_i$.

The expected FPR, which is constrained to be at most $F$ in our optimization problem, can be computed in the same manner using the calibration data.
However, as we will explain in the next section, by configuring the Bloom filter FPRs in a manner similar to PLBF~\cite{vaidya2021partitioned}, the expected FPR automatically matches the target value $F$.
Therefore, we defer a detailed discussion to the next section.

\section{CLBF Configuration Optimization}
\label{sec:clbf_configuration_optimization}

Our goal is to determine the optimal configuration parameters $D$, $\mathcal{F}$, and $\bm{\theta}$ that minimize the objective function in \cref{eq:objective}.
As shown in the \textbf{while} loop of \cref{alg:clbf_construction}, we adopt an alternating optimization framework.
For each candidate intermediate threshold vector $\bm{\theta}$ generated in the outer loop (Line~\ref{line:getnexttheta}), we solve an inner optimization problem to determine the optimal $D$ and $\mathcal{F}$ (Line~\ref{line:optdf}).
This inner optimization is carried out using dynamic programming, as described in \cref{sec:dynamic_programming_for_optimizing_d_and_f}.
The outer optimization adjusts $\bm{\theta}$ by repeatedly invoking the inner optimization, as detailed in \cref{sec:optimizing_intermediate_thresholds}.

\subsection{Dynamic Programming for \texorpdfstring{$D$ and $\mathcal{F}$}{D and F}}
\label{sec:dynamic_programming_for_optimizing_d_and_f}

We propose a dynamic programming-based method to compute the optimal $D$ and $\mathcal{F}$ for a fixed intermediate threshold vector $\bm{\theta}$ (Line~\ref{line:optdf} of \cref{alg:clbf_construction}).
Throughout this subsection, we assume that $\bm{\theta}$ is fixed and therefore omit $\bm{\theta}$ from the notation whenever it is clear from the context;
e.g., $\mathcal{L}(D,\mathcal{F})$ denotes $\mathcal{L}(D,\mathcal{F},\bm{\theta})$.

\textbf{Key Insight.}
The core observation enabling an efficient optimization is that once the product of the FPRs of upstream Trunk Bloom filters is fixed, the optimal configuration of all downstream Bloom filters is uniquely determined.
This lets us decompose the global optimization into overlapping subproblems indexed by the cascade depth and the upstream FPR product.

\begin{theorem_box}
\begin{theorem}[Sufficiency of Upstream FPR Product]
    \label{thm:sufficiency}
    Fix a cascade depth index $d \in \{1,\dots,D\}$.
    Given the product of the FPRs of the upstream $\mathrm{TBF}$s (i.e., $\prod_{j=1}^{d} f^{(\mathrm{t})}_j$),
    the optimal Bloom-filter FPRs at depth $d$ (i.e., $f^{(\mathrm{b})}_d$ if $D > d$, or $\{f^{(\mathrm{f})}_k\}_{k=1}^{K}$ if $D = d$) are uniquely determined.
\end{theorem}
\end{theorem_box}
\begin{proof}
First, by \cref{lem:query_flow}, the expected proportions of keys and non-keys entering the Bloom filters at depth $d$ are determined by the product of the FPRs of the upstream $\mathrm{TBF}$s:
they are $g^{(\mathrm{f})}_{D,k}$ and $h^{(\mathrm{f})}_{D,k} \prod_{j=1}^{D} f^{(\mathrm{t})}_j$ for $\mathrm{FBF}_k$ (if $D = d$), and $g^{(\mathrm{b})}_{d}$ and $h^{(\mathrm{b})}_{d} \prod_{j=1}^{d} f^{(\mathrm{t})}_j$ for $\mathrm{BBF}_d$ (if $D > d$).
Second, given these proportions, we can choose its optimal FPR in the same manner as in the PLBF paper~\cite[Sec.~3.3.1]{vaidya2021partitioned}.
Specifically, if these proportions are $g$ and $h$, respectively, then the optimal assignment is to set its FPR to $Fg/h$, where $F$ is the overall target FPR.
Therefore, if $d = D$, the optimal FPR of $\mathrm{FBF}_k$ is $(F g^{(\mathrm{f})}_{D,k}) / (h^{(\mathrm{f})}_{D,k} \prod_{j=1}^{D} f^{(\mathrm{t})}_j)$, and if $d < D$, the optimal FPR of $\mathrm{BBF}_d$ is $(F g^{(\mathrm{b})}_{d}) / (h^{(\mathrm{b})}_{d} \prod_{j=1}^{d} f^{(\mathrm{t})}_j)$.
This assignment ensures that the overall expected FPR equals $F$.
\end{proof}

\textbf{Definition of the $\mathrm{dp}$.}
Motivated by \cref{thm:sufficiency}, we define the DP value function $\mathrm{dp}(d, T)$ as follows:
\begin{equation}
    \label{eq:dp}
    \mathrm{dp}(d, T) \coloneq \min_{D, \mathcal{F}} \mathcal{L}_d (D,\mathcal{F}) \;\; \text{s.t.} \;\; \prod_{j=1}^{d-1} f^{(\mathrm{t})}_j = T,
\end{equation}
where $\mathcal{L}_d(D,\mathcal{F})$ denotes the objective restricted to the part of the cascade \textit{under} depth $d$.
Concretely, $\mathcal{L}_d(D,\mathcal{F})$ is defined analogously to $\mathcal{L}(D,\mathcal{F})$, except that all summations over the cascade depth index $i$ in \cref{eq:memory_size,eq:reject_time} start from $d$ instead of $1$; due to space constraints, we omit the explicit formulas here and refer the reader to \cref{sec:appendix:L_d_def}.
By definition, $\mathcal{L}_1(D,\mathcal{F}) = \mathcal{L}(D,\mathcal{F})$, and $\mathrm{dp}(1,1)$ corresponds to the original optimization problem.

Optimizing $f^{(\mathrm{t})}_d$ over the continuous range $(0,1]$ is computationally infeasible, so we approximate the optimization by discretizing the domain:
we restrict $f^{(\mathrm{t})}_d$ and $T$ to a finite set $\mathcal{P} = \{1, 0.5, 0.5^2 \dots, 0.5^{P-1}\}$, where $P \in \mathbb{N}$ (in our experiment, we set $P=20$).
This discretization is reasonable and natural because, under the standard Bloom filter design that maintains a bit density close to $0.5$, a Bloom filter with $k \in \mathbb{Z}_{\ge 0}$ hash functions has a FPR approximately equal to $0.5^{k}$.
Moreover, the discretization is sufficiently fine: when $P=20$, the smallest value is $0.5^{P-1}\approx 2\times10^{-6}$, which is small enough for practical purposes.
This natural discretization allows $\mathrm{dp}(d, T)$ to be treated as a computationally tractable dynamic programming table with $\bar{D} \times P$ states.

\begin{figure}[t]
    \centering
    \begin{minipage}{0.23\textwidth}
        \centering
        \includegraphics[width=\textwidth]{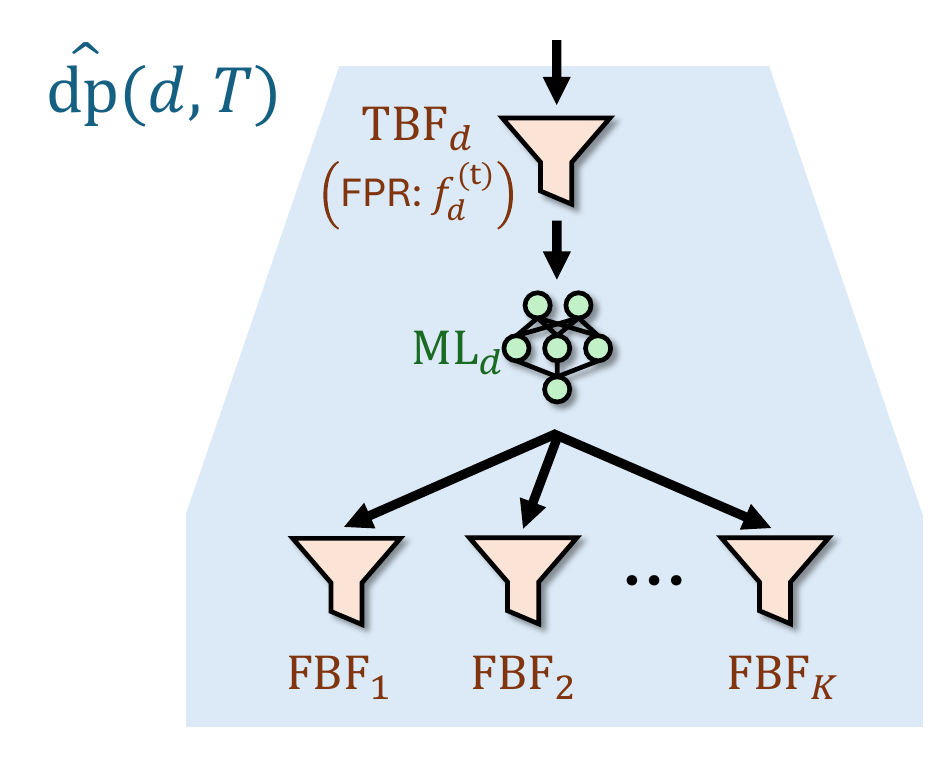}
        \subcaption{$\hat{\mathrm{dp}}(d, T)$ is the minimum objective function value under $\mathrm{TBF}_d$ subject to the constraint that $\prod_{j=1}^{d-1} f^{(\mathrm{t})}_j = T$ and $D = d$.}
        \label{fig:dp1}
    \end{minipage}
    \hfill
    \begin{minipage}{0.23\textwidth}
        \centering
        \includegraphics[width=\textwidth]{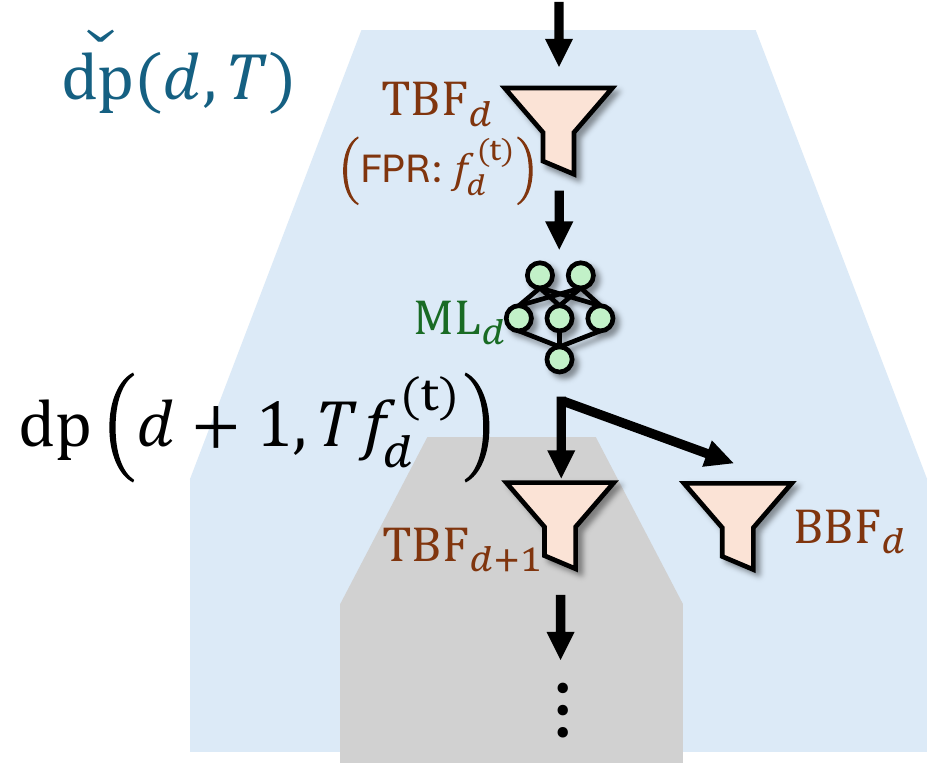}
        \subcaption{$\check{\mathrm{dp}}(d, T)$ is the minimum objective function value under $\mathrm{TBF}_d$ subject to the constraint that $\prod_{j=1}^{d-1} f^{(\mathrm{t})}_j = T$ and $D > d$.}
        \label{fig:dp2}
    \end{minipage}
    \caption{The value of $\mathrm{dp}(d,T)$ is calculated by selecting the appropriate value from the case where $D=d$, i.e., $\hat{\mathrm{dp}}(d, T)$, and the case where $D>d$, i.e., $\check{\mathrm{dp}}(d, T)$.}
    \Description{Illustration of the dynamic programming formulation for CLBF optimization, showing two cases at each depth: (a) terminating with final Bloom filters or (b) continuing the cascade with branching.}
\end{figure}

\textbf{Calculation of $\mathrm{dp}(d, T)$.}
We calculate $\mathrm{dp}(d, T)$ recursively by considering whether the cascade terminates at the current depth $d$ or continues to a deeper level.
In the first case, $D=d$, $\mathrm{ML}_d$ is the last ML model in the cascade, and the queries leaving $\mathrm{ML}_d$ are routed directly to the $\mathrm{FBF}$s (\cref{fig:dp1}).
In the second case, $D>d$, $\mathrm{ML}_d$ is followed by another ML stage: two-way branching is introduced below $\mathrm{ML}_d$, and the subtree that continues to the next depth is optimized recursively (\cref{fig:dp2}).
We evaluate the best configuration in each case and select the one with the smaller objective value.

We define $\hat{\mathrm{dp}}(d, T)$ as the solution to the optimization problem (\cref{eq:dp}) under the additional constraint $D=d$.
Similarly, we define $\check{\mathrm{dp}}(d, T)$ as the solution under the additional constraint $D>d$.
Then, $\mathrm{dp}(d, T)$ is given by:
\begin{equation}
    \label{eq:dp_recursive}
    \mathrm{dp}(d, T) = \begin{dcases}
        \hat{\mathrm{dp}}(d, T) & (d = \bar{D}), \\
        \min\left(\hat{\mathrm{dp}}(d, T), \check{\mathrm{dp}}(d, T)\right) & \text{(else)}.
    \end{dcases}
\end{equation}
Note that when $d=\bar{D}$, $\check{\mathrm{dp}}(d, T)$ is not evaluated, because the cascade cannot contain an additional ML stage.

\textbf{Calculation of $\hat{\mathrm{dp}}(d, T)$.}
We can compute the value of $\hat{\mathrm{dp}}(d, T)$ because $T = \prod_{j=1}^{d-1} f^{(\mathrm{t})}_j$ encodes all the necessary information about the upstream $\mathrm{TBF}$s and therefore fully determines the configuration at depth $d$ (\cref{fig:dp1}).
Once we fix $f^{(\mathrm{t})}_d \in \mathcal{P}$, the optimal FPR of each $\mathrm{FBF}$ can be determined by following the proof of \cref{thm:sufficiency}.
For each candidate $f^{(\mathrm{t})}_d$, we evaluate the objective function in \cref{eq:dp} and select the smallest objective value to determine $\hat{\mathrm{dp}}(d, T)$.

\textbf{Calculation of $\check{\mathrm{dp}}(d, T)$.}
We can calculate $\check{\mathrm{dp}}(d, T)$ in a similar way: once we fix $f^{(\mathrm{t})}_d \in \mathcal{P}$, the optimal FPR for $\mathrm{BBF}_d$ can be determined by following the proof of \cref{thm:sufficiency}.
For the remaining part of the cascade, starting from $\mathrm{TBF}_{d+1}$ (i.e., the gray shaded area in \cref{fig:dp2}), the optimal configuration can be obtained by evaluating $\mathrm{dp}(d+1, T f^{(\mathrm{t})}_d)$.
This is because, by definition, $\mathrm{dp}(d+1, T f^{(\mathrm{t})}_d)$ gives the minimum value of the objective function when the structure from depth $d+1$ to $D$ is optimally configured under the constraint $\prod_{j=1}^{d} f^{(\mathrm{t})}_j = T f^{(\mathrm{t})}_d$.
Since we assume $T=\prod_{j=1}^{d-1} f^{(\mathrm{t})}_j$, this constraint is satisfied, enabling this recursive evaluation.

\textbf{Computational Complexity.}
Without dynamic programming, computing the optimal configuration under discretization would require exhaustive enumeration over the FPRs of TBFs.
Since each of the $\bar{D}$ TBFs takes one of $P$ discrete values, the cost is $\mathcal{O}(P^{\bar{D}})$, which makes the computation intractable.

Our dynamic program reduces this to polynomial time.
A naive implementation runs in $\mathcal{O}(\bar{D}P^2K)$.
There are $\bar{D}P$ states $(d,T)$.
For each state, evaluating $\hat{\mathrm{dp}}(d,T)$ costs $\mathcal{O}(PK)$: we try $P$ candidates for $f^{(\mathrm{t})}_d$ and, for each, compute the optimal rates for all $K$ $\mathrm{FBF}$s.
Evaluating $\check{\mathrm{dp}}(d,T)$ costs $\mathcal{O}(P)$: for each candidate $f^{(\mathrm{t})}_d$, we optimize one $\mathrm{BBF}$ and look up $\mathrm{dp}(d+1, Tf^{(\mathrm{t})}_d)$. Thus the total is $\mathcal{O}(\bar{D}P \cdot PK)+\mathcal{O}(\bar{D}P\cdot P)=\mathcal{O}(\bar{D}P^2K)$.

We can reduce the time complexity to $\mathcal{O}(\bar{D}PK + \bar{D}P^2)$ by caching the $\mathrm{FBF}$ computations.
The optimal $\mathrm{FBF}$ configuration depends only on $Tf^{(\mathrm{t})}_d$.
Hence, for all $\mathcal{O}(\bar{D}P)$ distinct pairs $(d, Tf^{(\mathrm{t})}_d)$, we precompute the optimal $\mathrm{FBF}$ rates and their total memory in $\mathcal{O}(\bar{D}PK)$ time.
Afterwards, each evaluation of $\hat{\mathrm{dp}}(d,T)$ takes only $\mathcal{O}(P)$ time, yielding a total complexity of $\mathcal{O}(\bar{D}PK+\bar{D}P^2)$.
With $\bar{D}\approx 100$, $P\approx 20$, and $K\approx 10$, this cost is sufficiently small in practice.

\subsection{Optimizing Intermediate Thresholds \texorpdfstring{$\bm{\theta}$}{theta}}
\label{sec:optimizing_intermediate_thresholds}

We generate candidates for the intermediate thresholds $\bm{\theta}$ using $\mathrm{GetNextTheta}$ (Line~\ref{line:getnexttheta} of \cref{alg:clbf_construction}) and iterate until no further candidate is returned (Line~\ref{line:ifthetaisnonebreak}).
We implement $\mathrm{GetNextTheta}$ using two complementary strategies.

The first strategy enumerates quantile-based threshold candidates.
For each $\alpha \in \{0.1, 0.01, 0.001, 0.0001, 0.0\}$, we construct $\bm{\theta}$ by setting each $\theta_d$ to the upper $\alpha$-quantile of the non-key scores in the calibration set produced by $\mathrm{ML}_d$ ($d \in \{1,2,\dots,\bar{D}-1\}$).
This quantile-based approach allows each layer to account for the score distribution of its corresponding ML model.

The second strategy further refines the intermediate thresholds using Bayesian optimization.
We warm-start the optimizer with the quantile-based candidates, and then run Bayesian optimization for 100 iterations to further reduce the objective.

\section{Experiments}
\label{sec:experiments}

All experiments were conducted on a Linux machine equipped with an Intel\textsuperscript{\textregistered}~Core\texttrademark{}~i9-11900H CPU @ 2.50\,GHz and 64\,GB of memory.
While the majority of the system was implemented in C++, two components were executed in Python: the Bayesian optimization of threshold parameters using Optuna~\cite{akiba2019optuna} and the original implementation of PHBF~\cite{bhattacharya2022new}.
The C++ code was compiled using GCC version 11.4.0 with the \texttt{-O3} optimization flag, and all experiments were conducted in single-threaded mode.

\textbf{Key takeaways:} The main findings are as follows:
\begin{itemize}[leftmargin=1.5em]
    \item CLBF achieves a superior trade-off between memory and FPR, reducing memory usage by up to 17\% over PLBF (the most memory-efficient LBF under a fixed machine learning model).
    \item CLBF achieves a superior trade-off between memory and reject time, reducing reject time by up to 14.6$\times$ compared to PLBF under our default XGBoost setting.
    \item Although CLBF solves a more complex optimization problem within a unifying framework that generalizes existing LBFs, it maintains a construction time comparable to existing LBFs, ranging from 1.003$\times$ to 1.23$\times$ longer.
    \item An ablation study varying the ML model size shows that, while the memory efficiency of existing LBFs strongly depends on the given model size, CLBF preserves memory efficiency by automatically reducing the model to an appropriate size.
    \item In the appendix, we additionally evaluate alternative ML architectures and XGBoost hyperparameters; CLBF consistently achieves superior memory--FPR and memory--reject-time trade-offs across these variants, with reject-time improvements of up to 65$\times$ compared to PLBF.
    \item The optimization of intermediate thresholds using Optuna yields an additional improvement over the quantile-based candidate selection method in memory efficiency and reject time, at the cost of a substantially longer construction time.
    \item CLBF achieves a memory--FPR trade-off comparable to Gama, while substantially outperforming it in the memory--reject-time trade-off (by up to 25$\times$) and construction time (by up to 6$\times$).
\end{itemize}

\subsection{Compared Methods}
We compare CLBF with the following baselines:
\begin{itemize}[leftmargin=1.5em]
    \item \textbf{Bloom Filter}~\cite{bloom1970space}: We used a public C++ implementation~\cite{github_fastfilter_cpp}.
    \item \textbf{PLBF}~\cite{vaidya2021partitioned}: We implemented PLBF in C++ with $N = 100$ segments and $K = 8$ regions. We used Fast PLBF~\cite{sato2023fast}, which builds exactly the same data structure as PLBF.
    \item \textbf{Sandwiched LBF}~\cite{mitzenmacher2018model}: We implemented Sandwiched LBF in C++. We select the threshold by evaluating $N=100$ candidate thresholds and selecting the best one.
    \item \textbf{G-PLBF}~\cite{shang2025gama}: We implemented G-PLBF, i.e., PLBF optimized by Gama~\cite{shang2025gama}, in C++ with $N = 100$ segments and $K = 8$ regions. We used Fast PLBF~\cite{sato2023fast}. 
    \item \textbf{Ensemble LBF (ELBF)}~\cite{lin2025ensemble}: We implemented ELBF in C++, using the same hyperparameter settings as in the original paper (i.e., $\Delta=0.1, \epsilon=0.01, \delta=0.01$). We swept $T$, i.e., the total space budget, to obtain different operating points.
    \item \textbf{Projection Hash Bloom Filter (PHBF)}~\cite{bhattacharya2022new}: We evaluated two implementations: \textbf{PHBF~(Python)}, the authors' original Python code~\cite{githubbhattacharya2022new}, and \textbf{PHBF~(C++)}, our optimized C++ reimplementation. We set the sampling factor $s = 100$, number of hash functions $k = 30$.
    \item \textbf{Hash Adaptive Bloom Filter (HABF)}~\cite{xie2021hashadaptive,li2023pareto}: We used the author's C++ implementation~\cite{githubnjulands2021hashadaptivebf}. We vary the hyperparameter $\texttt{bits\_per\_key}$ from $1$ to $15$.
\end{itemize}
Unlike CLBF, PLBF and Sandwiched LBF do not automatically reduce ML model size; thus, we evaluate multiple model sizes for them.
We did not include Ada-BF and Disjoint Ada-BF~\cite{dai2020adaptive} as baselines because they essentially share the same structure as PLBF, while PLBF is known to be strictly superior in terms of memory usage and accuracy due to its more thorough parameter optimization.

We evaluate two variants of our method:
\textbf{CLBF}, in which the intermediate thresholds $\bm{\theta}$ are selected from an adaptively defined candidate set,
and \textbf{CLBF w/ Optuna}, in which $\bm{\theta}$ is further refined via Bayesian optimization using Optuna~\cite{akiba2019optuna} (as described in \cref{sec:optimizing_intermediate_thresholds}).
CLBF w/ Optuna incurs a substantially longer construction time but provides modest performance improvements.

While any AMQ can be used inside LBFs (including CLBF), for fairness we use a standard Bloom filter in all experiments, and accordingly as the non-learned baseline.

\subsection{Machine Learning Models}
\label{sec:exp:ML models}
In the main experiments below, we use \textbf{XGBoost} as the ML model for Sandwiched LBF, PLBF, and CLBF.
Each weak learner in XGBoost corresponds to each ML model $\mathrm{ML}_1, \dots, \mathrm{ML}_{\bar{D}}$.
We set \texttt{num\_round} ($=\bar{D}$) to $100$ and \texttt{max\_depth} to $4$.
Prior work~\citep{dai2022optimizing} suggests that tree-based ensemble models are strong and reliable candidates for the ML component in LBFs, and we also observe that XGBoost achieves stable performance across memory usage, accuracy, and reject time.

We additionally evaluate alternative ML architectures and XGBoost hyperparameter settings in \cref{sec:appendix:ml_model_variants}.

\subsection{Datasets}

\begin{table}[t]
    \centering
    \caption{Statistics of the datasets.}
    \label{tab:datasets}
    \setlength{\tabcolsep}{4pt}
    \begin{tabular}{@{}lrrrr@{}}
        \toprule
        Dataset & dim. & $|\mathcal{S}|$ & $|\mathcal{Q}|$ & \#non-key (test)\\
        \midrule
        URL & 20 & 223{,}088 & 418{,}103 & 10{,}000 \\
        Ember & 2{,}381 & 400{,}000 & 390{,}000 & 10{,}000 \\
        Higgs & 28 & 5{,}829{,}123 & 5{,}160{,}877 & 10{,}000 \\
        \bottomrule
    \end{tabular}
\end{table}

We used the following three datasets:
\begin{itemize}[leftmargin=1.5em]
    \item \textbf{URL}~\cite{manu2021urlDataset}, containing malicious URLs (keys) and benign URLs (non-keys).
    We extracted 20 lexical features, e.g., URL length and number of special characters.
    \item \textbf{Ember}~\cite{anderson2018ember}, containing malicious files (keys) and benign files (non-keys), with $2{,}381$-dimensional features and sha256 hashes. We did not use unlabeled files.
    \item \textbf{Higgs}~\cite{baldi2014higgs}, a dataset of $28$-dimensional features for distinguishing Higgs signals from background.
    We use label $1$ (signal) as keys and label $0$ (background) as non-keys.
\end{itemize}
\cref{tab:datasets} shows the sizes of $\mathcal{S}$ and $\mathcal{Q}$, as well as the feature dimension.
We use all keys in the dataset as $\mathcal{S}$, and define $\mathcal{Q}$ as the set of non-keys after removing the $10{,}000$ test non-keys.
All keys are used for training and calibration.
Among $\mathcal{Q}$, we allocate $90\%$ to the training set and $10\%$ to the calibration set.

For every dataset, we use $10{,}000$ non-keys as test queries, which are never used for training or calibration.
This differs from prior work such as HABF~\citep{xie2021hashadaptive,li2023pareto} and PHBF~\citep{bhattacharya2022new}, where non-keys used in the training data are also used for testing.
In our experiments, the data structure is always evaluated on unseen non-keys.
For brevity, we present detailed results in the main text on three representative datasets (URL, Ember, and Higgs);
comprehensive experiments on 14 datasets are reported in \cref{sec:appendix:comprehensive}.

\subsection{Evaluation Metrics}
In our experiments, we evaluate four metrics: memory usage, FPR, reject time, and construction time.

\textbf{Memory usage} is the total memory consumed by the data structure.
For LBFs, it includes both the Bloom-filter components and the underlying ML model.
For PHBF, it includes both the bit array and the projection vectors, following the original PHBF paper~\cite{bhattacharya2022new}.

\textbf{FPR} is the probability that the data structure returns a positive answer for a non-key query.
We evaluate it on the test data (i.e., the $10{,}000$ non-keys) by computing the fraction of test non-keys that are incorrectly accepted (i.e., classified as positive).

\textbf{Reject time} is the time required for the data structure to process a non-key query and return an answer.
We measure it by averaging the query time over all test non-keys.
Note that this is an end-to-end time: for LBFs, the reject time includes the ML inference overhead as well as any additional operations within the data structure.

\textbf{Construction time} is the end-to-end time to build the data structure, consisting of the following four components:
\textbf{(i) ML Model Training}: the time to train the ML model.
\textbf{(ii) Calibration}: the time to run inference on the calibration data to obtain model scores.
\textbf{(iii) Configuration}: the time to compute the optimal configuration (including the time for $\mathrm{BenchmarkClassicBF}$).
\textbf{(iv) Add Keys}: the time to insert the keys into the data structure.

\begin{figure}[t]
    \centering
    \includegraphics[width=\columnwidth]{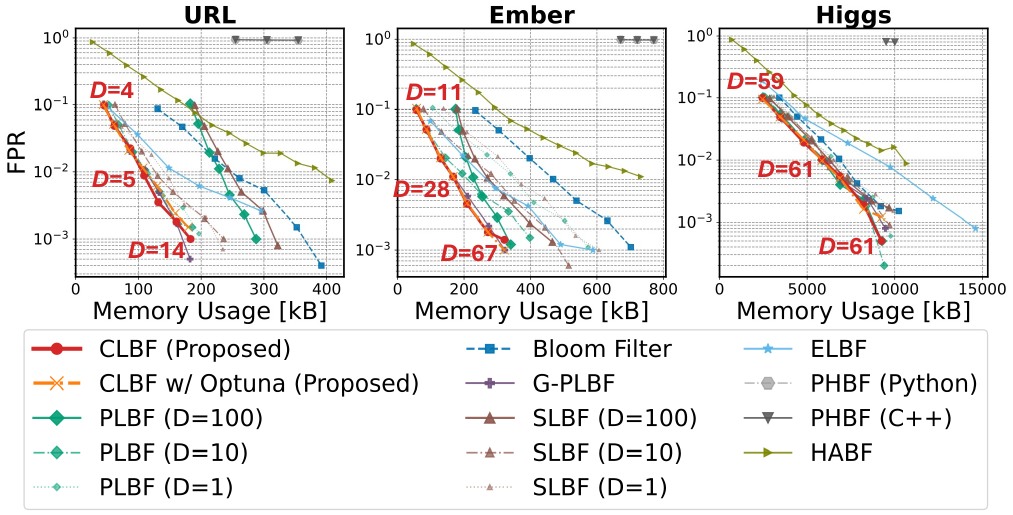}
    \caption{Trade-off between memory usage and FPR (lower-left is better):
    CLBF consistently matches or surpasses existing LBFs, achieving up to 17\% less memory usage than the best PLBF configuration at the same FPR.}
    \label{fig:exp:memory_accuracy_trade_off}
    \Description{Trade-off between memory usage and FPR across datasets, showing that CLBF achieves better memory efficiency than existing LBFs.}
\end{figure}

\begin{figure}[t]
    \centering
    \includegraphics[width=\columnwidth]{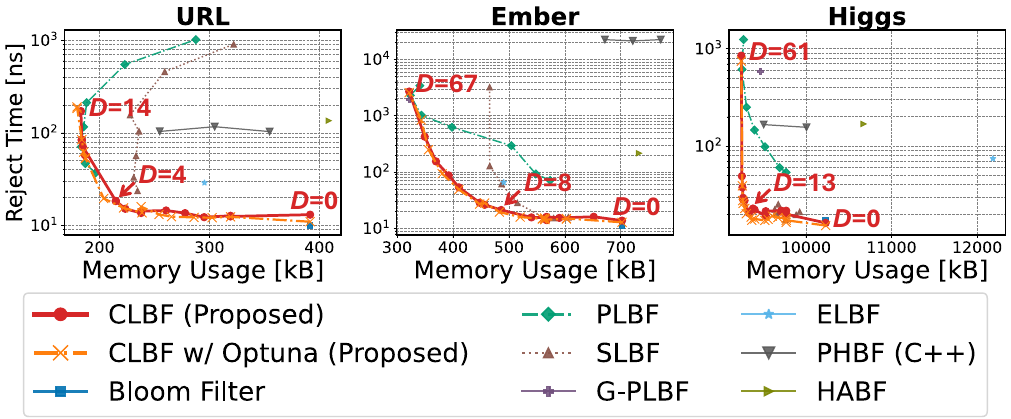}
    \caption{Trade-off between memory usage and average reject time (lower-left is better):
    CLBF achieves equal or lower reject time than existing LBFs at the same memory budget, with up to 14.6$\times$ faster rejection than PLBF.}
    \label{fig:exp:memory_reject_time_trade_off}
    \Description{Trade-off between memory usage and average reject time across datasets, demonstrating that CLBF achieves faster rejection than existing LBFs at similar memory budgets.}
\end{figure}

\subsection{Main Results}
\label{sec:exp:main_results}
\textbf{Memory vs FPR.}
We evaluate the memory--FPR trade-off by varying the target FPR $F$ from 0.1 to 0.001.
For CLBF, we set $\lambda=1.0$ (optimizing solely for memory) and the number of boosting rounds $\bar{D}$ to 100.
For ELBF, we use the $\bar{D}=100$ decision trees as the oracle pool.
Since other LBFs cannot automatically adjust the ML model size, we report results for $D\in\{1, 10, 100\}$.

\cref{fig:exp:memory_accuracy_trade_off} illustrates the results.
For CLBF at $F\in\{0.1, 0.01, 0.001\}$, we annotate the selected value of $D$.
We observe that CLBF consistently achieves equal or better memory efficiency than existing baselines at the same FPR.
For example, on the Ember dataset with $F = 0.005$, CLBF reduces memory usage by 17\% compared to the best configuration of PLBF.

Moreover, the optimal size of the ML model depends on the dataset and the target FPR, and CLBF automatically adapts to this.
On the Ember and Higgs datasets, a medium-sized model ($D \approx 20$--$60$) is frequently optimal, whereas on the URL dataset, where the binary classification problem is relatively easy, a smaller model with $D$ around $10$ is sufficient.
We also observe that the selected $D$ tends to increase as the target FPR decreases.
Bayesian optimization of the intermediate thresholds can marginally reduce memory usage in some settings.
Unlike PLBF, CLBF achieves such strong memory efficiency without manually trying multiple values of $D$.
G-PLBF achieves memory usage comparable to that of CLBF, as both methods improve overall memory efficiency by adjusting the allocation between the ML model and the filters.

All baselines except G-PLBF exhibit worse memory efficiency than PLBF and thus also than CLBF.
Sandwiched LBF is less memory-efficient than PLBF because it does not exploit the scores produced by the ML model as effectively.
In some settings, ELBF achieves better memory efficiency than PLBF configured with an inappropriately large ML model (e.g., $D = 100$), since it can prune the oracle ensemble and avoid an unnecessarily large one.
On the other hand, ELBF can also be less memory-efficient than PLBF, because it uses fixed thresholds without a mechanism to optimize them as in PLBF and CLBF, and its objective assumes independent oracles, which may not hold in practice.
PHBF shows very high FPR (often near $1$) on several datasets, likely because its projection-based hashing tends to collide when the feature vector of a non-key query is similar to that of any stored key.
Increasing the number of hashes $k$ may reduce such collisions, but requires storing more projection vectors and thus increases memory.
HABF also underperforms in our setting because it targets a different regime: it minimizes FPR for a fixed set of non-keys known at construction time, whereas standard Bloom filters and LBFs are better when the set of non-keys is not fixed in advance.

\begin{figure}[t]
    \centering
    \includegraphics[width=\columnwidth]{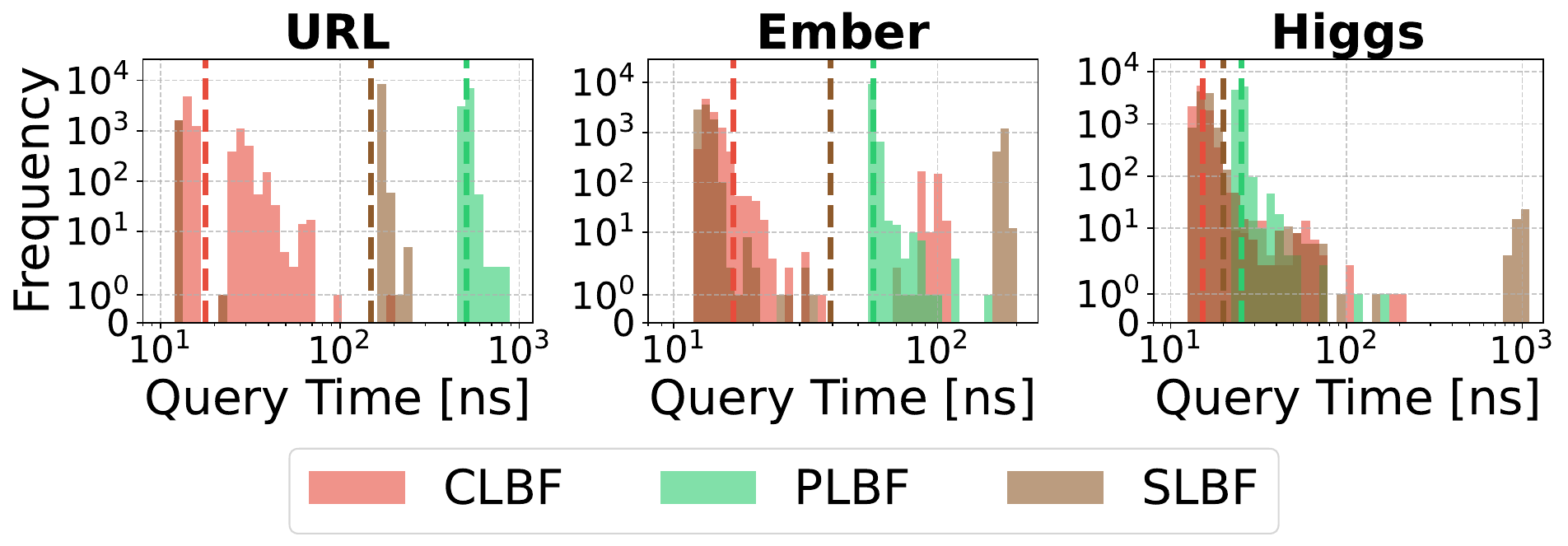}
    \caption{Distribution of reject time:
    CLBF returns most queries quickly via intermediate Bloom filters and early exit, whereas PLBF and Sandwiched LBF incur full ML inference for all or some queries.}
    \label{fig:exp:query_time_histogram}
    \Description{Histogram of reject time distribution, showing that CLBF achieves fast rejection for most queries through intermediate filters and early exit, while PLBF and Sandwiched LBF have different characteristics.}
\end{figure}

\textbf{Memory vs Reject Time.}
We evaluate the trade-off between memory usage and reject time, which is controlled in CLBF by varying the hyperparameter $\lambda$ within $[0, 1]$.
We set the FPR $F$ to 0.001 for all methods and the number of XGBoost boosting rounds $\bar{D}$ to 100 for CLBF.
Since the standard Bloom filter lacks a parameter to control this trade-off, it is represented as a point.
For Sandwiched LBF and PLBF, we vary $D$ from 1 to 100 to observe changes in memory usage and reject time.
For ELBF, we report the configuration that achieves the FPR closest to $0.001$ from the memory--FPR experiment (\cref{fig:exp:memory_accuracy_trade_off}).
For PHBF and HABF, we were unable to obtain a case with a sufficiently small FPR, so we show the results only as a reference.
Note that it is \textbf{not} a fair comparison because the FPR is completely different between PHBF/HABF and the other methods.
We omit PHBF (Python) from plots because PHBF (C++) consistently achieves shorter reject time across all datasets.

\cref{fig:exp:memory_reject_time_trade_off} shows the results.
For CLBF at $\lambda\in\{0.0, 0.9, 1.0\}$, we annotate the selected value of $D$.
CLBF greatly outperforms Sandwiched LBF in terms of memory efficiency and PLBF in terms of reject time.
While Sandwiched LBF has shorter reject time than PLBF, its memory efficiency is worse than both: for example, on the Ember dataset, CLBF achieves a minimum memory usage of $321\,\mathrm{kB}$, whereas no Sandwiched LBF configuration uses less than $465\,\mathrm{kB}$.
On the other hand, although PLBF can achieve memory efficiency comparable to CLBF, it incurs significantly longer reject times.
Specifically, on the Higgs dataset, when the memory usage is around $9{,}250\,\mathrm{kB}$ (a budget unattainable by Sandwiched LBF), the average reject time of CLBF is approximately 14.6$\times$ shorter than that of PLBF.

Moreover, when $\lambda=0.0$, we have $D=0$ and CLBF reduces to a classical AMQ filter (in our experiments, a Bloom filter); varying $\lambda$ from $0$ to $1$ thus continuously interpolates between a reject-time-optimal Bloom filter and a memory-optimal LBF.
ELBF and G-PLBF exhibit longer reject times than CLBF because they are not designed to minimize reject time.
However, they often achieve shorter reject times than PLBF and Sandwiched LBF configured with an inappropriately large ML model (e.g., $D=100$), since ELBF prunes the oracle ensemble and G-PLBF optimizes memory allocation, both avoiding an unnecessarily large ensemble.
Additionally, we find that threshold optimization via Bayesian optimization yields slight improvements in reject time.
This confirms that selecting from an adaptively defined set of threshold candidates is a fast and reasonably effective approach, and suggests that further optimization of intermediate thresholds could reduce reject time even more.

\cref{fig:exp:query_time_histogram} shows the distribution of reject time for each LBF.
We set the target FPR to $F=0.001$ and choose configurations with roughly matched memory usage per dataset: about $220\,\mathrm{kB}$ on URL, $460\,\mathrm{kB}$ on Ember, and $9{,}700\,\mathrm{kB}$ on Higgs.
Both axes are in logarithmic scale, and the dashed line indicates the mean reject time.
CLBF returns most queries quickly, benefiting from early rejection by intermediate Bloom filters and early exit via branching.
In contrast, PLBF performs full ML model inference for every query, resulting in consistently longer reject times.
Sandwiched LBF achieves fast rejection for many queries, similar to CLBF; however, queries that pass through the pre-filter require full ML model inference, which for some queries yields substantially longer reject times than CLBF.

\begin{figure*}[t]
    \centering
    \includegraphics[width=\textwidth]{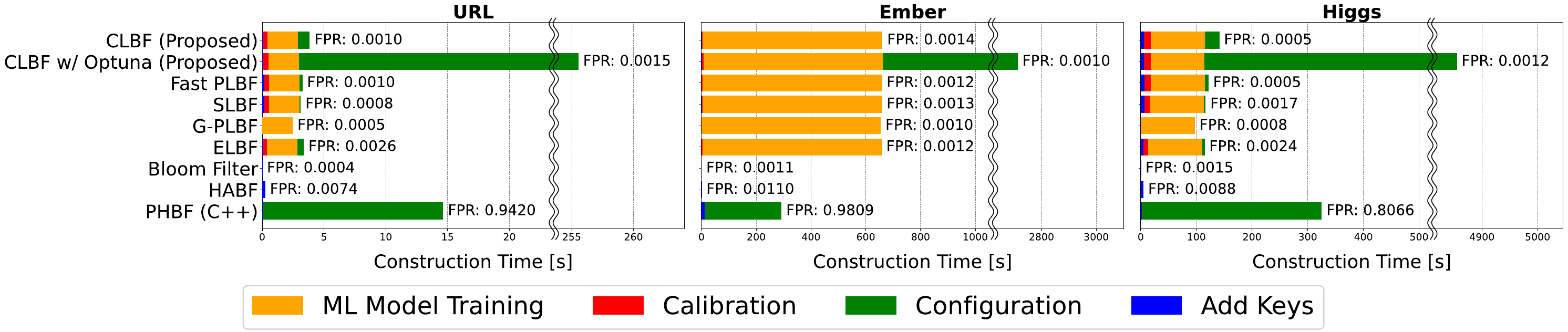}
    \caption{Construction time: CLBF introduces only a minor additional overhead, requiring at most 1.23$\times$ the construction time of existing LBFs despite searching for a near-optimal configuration.}
    \label{fig:exp:time_hist}
    \Description{Construction time comparison showing that CLBF incurs only a small overhead compared to existing LBFs despite additional optimization.}
\end{figure*}

\begin{figure}[t]
    \centering
    \includegraphics[width=\columnwidth]{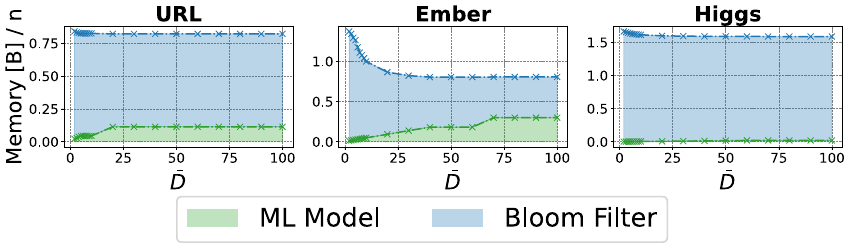}
    \caption{Effect of the number of given models $\bar{D}$ in CLBF: Increasing $\bar{D}$ reduces memory usage until it becomes sufficient, after which CLBF selects a smaller optimal subset $D$ and the memory usage remains unchanged.}
    \label{fig:exp:D_memory_clbf}
    \Description{As the number of available models increases, CLBF reduces memory usage until enough models are provided, after which it selects a smaller optimal subset and memory usage remains stable.}
\end{figure}

\begin{figure}[t]
    \centering
    \includegraphics[width=\columnwidth]{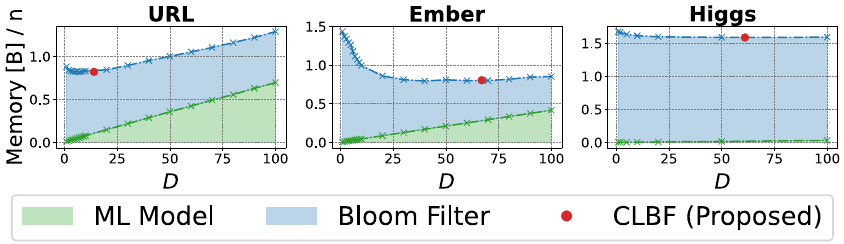}
    \caption{Effect of the number of models $D$ in PLBF: Increasing $D$ initially reduces the total memory usage, but eventually increases it because PLBF must retain all $D$ models.}
    \label{fig:exp:D_memory_plbf}
    \Description{As the number of models increases, PLBF first reduces memory usage, but eventually uses more memory because all models must be kept.}
\end{figure}

\subsection{Construction Time}
We compare the construction time of CLBF with the baselines.
For the LBFs, the number of weak learners is set to $100$.
Again, note that HABF and PHBF do not achieve FPRs comparable to the other baselines, so the comparison is \textbf{not} entirely fair.
We omit the results of PHBF (Python) from the plots, as it is consistently slower than PHBF (C++).

The results are shown in \cref{fig:exp:time_hist}.
The construction time of CLBF is 1.003$\times$ to 1.23$\times$ longer than that of Sandwiched LBF.
This overhead mainly comes from determining the optimal structure: CLBF spends up to 16$\times$ longer on this step than Sandwiched LBF, which selects its configuration using a simpler algorithm without dynamic programming.
However, this step accounts for only 0.3\% to 24\% of the total construction time, so its impact is marginal relative to CLBF's gains in memory efficiency and reject time.

Even Sandwiched LBF already requires construction times 180 to 1{,}100{,}000$\times$ longer than those of a standard Bloom filter, reflecting different use cases: Bloom filters prioritize extremely fast construction, whereas LBFs invest more work at build time to reduce memory.
Thus, Bloom filters are preferable when rebuilds are frequent, while LBFs are preferable when memory is scarce and reconstruction is infrequent (e.g., malicious-URL filtering, where the key set typically changes slowly).
In such scenarios, CLBF's extra construction time is unlikely to be a practical issue, and the benefit of obtaining a near-optimal configuration by effectively searching over all $D \in \{1,2,\dots,100\}$ likely outweighs this cost.

The construction time of G-PLBF is substantially longer than that of CLBF.
G-PLBF relies on generic black-box optimization with sequential evaluation of candidate configurations, whereas CLBF leverages the cascade structure and performs a systematic search over a discretized configuration space via dynamic programming, thereby determining a configuration in much less time.
This suggests that structured optimization tailored to the proposed architecture can yield high-quality configurations while keeping construction time manageable.

Finally, while CLBF w/ Optuna can occasionally achieve slightly better memory efficiency or reject time than CLBF, it requires approximately 4.1 to 66.4$\times$ longer construction time.
Such a cost may be unacceptable in typical settings, but it can be justified when reconstruction is extremely infrequent or when a single optimized structure is deployed across many devices.
We therefore view CLBF as the more practical default choice, with CLBF w/ Optuna serving as an optional enhancement for highly specialized deployment scenarios.


\subsection{Ablation Study on Number of ML Models}
\label{sec:exp:ablation_study_on_D}

We experimentally show that our CLBF achieves an effective balance between model and filter sizes by appropriately reducing the learned model.

\cref{fig:exp:D_memory_clbf} shows the relationship between the number of ML models given to construct CLBF, i.e., $\bar{D}$, and the model--filter memory size balance chosen as a result of optimization (we always set $F=0.001$ here). 
When $\bar{D}$ is small, increasing $\bar{D}$ monotonically decreases the overall memory usage and monotonically increases the size of the ML model used.
On the other hand, for a certain range of large values of $\bar{D}$, the size of the Bloom filter and the ML models used do not change because CLBF selects an optimal subset $D$ from the given $\bar{D}$ models.

In contrast, in the case of PLBF~\cite{vaidya2021partitioned}, setting a larger $D$ can lead to an increase in memory usage.
The relationship between the number of ML models given when constructing PLBF, i.e., $D$, and the memory usage of PLBF is shown in \cref{fig:exp:D_memory_plbf} (we always set $F=0.001$ here).
As $D$ increases, of course, the memory usage of the ML models in PLBF increases, because PLBF uses all of the $D$ ML models given.
On the other hand, the memory usage of the backup Bloom Filter used by PLBF tends to decrease.
This is because the accuracy of the ML model tends to improve as the size of the ML model increases, and even a small Bloom Filter can achieve the target FPR.
The overall memory usage decreases in the range of a certain small $D$, and increases in the range of a certain large $D$.
The optimal $D$ varies depending on the dataset, but in all cases, CLBF (shown as a red dot) selects a $D$ close to the optimal value.

\section{Discussion}
\label{sec:discussion}

Having established the empirical effectiveness of CLBF in \cref{sec:experiments}, we now discuss its relationship to concurrent approaches, its current limitations, and promising directions for future work.

\textbf{Comparison with Gama~\cite{shang2025gama}.}
Independently and concurrently with our work, Shang et al.~\cite{shang2025gama} proposed Gama, a general framework for adaptively optimizing memory allocation between the ML model and filter components in learned Bloom filters.
The main difference between Gama and CLBF lies in the optimization formulation.
Gama treats configuration selection as a generic black-box optimization problem and primarily aims to improve FPR under a fixed memory budget.
In contrast, CLBF explicitly exploits the cascade structure and systematically searches a discretized configuration space, enabling joint optimization of model size and Bloom-filter parameters within the proposed architecture.

The difference is also reflected in the objective.
Whereas Gama focuses on memory--FPR trade-offs, CLBF additionally incorporates reject time into the formulation.
As shown in \cref{sec:experiments}, this distinction is practically important: under comparable memory budgets, CLBF achieves substantially faster rejection than G-PLBF.
This highlights that memory allocation in LBFs affects not only memory efficiency and FPR but also reject time, and that incorporating reject time into the problem formulation is important in practice.

\textbf{Further Optimization of Intermediate Thresholds.}
Optimizing intermediate thresholds remains a challenging open problem.
We tried two approaches: a simple candidate selection method and Bayesian optimization.
While the Bayesian approach improved CLBF performance, it incurred substantially higher computational cost, limiting its practicality.
More efficient and effective threshold-tuning methods could further improve the applicability of CLBF.

\textbf{Extension to Single Large Models.}
Additionally, although our method is well-suited for models composed of multiple weak learners (e.g., boosting), it cannot be directly applied to single large models (e.g., deep neural networks).
To enable their use in CLBF, a mechanism for extracting tentative scores from intermediate layers is required.
A key challenge is mitigating the additional memory and computational overhead introduced by this mechanism.
Addressing this issue may require new optimization strategies that carefully trade off early-exit accuracy, model footprint, and inference cost.

\textbf{Combination with ELBF~\cite{lin2025ensemble}.}
Our experiments show that CLBF achieves better memory--FPR and memory--reject-time trade-offs than ELBF under the authors' recommended hyperparameters.
Nevertheless, ELBF's core idea that optimizing an arbitrary \emph{subset} of candidate models rather than a \emph{prefix} of an ordered pool explores a richer model-selection space and may be advantageous in regimes where the best cascade is not prefix-structured.
Bringing this idea into CLBF is non-trivial: subset selection must be combined with cascade-specific execution paths and an explicit reject-time objective, which turns construction into a much harder combinatorial search problem.
Designing practical, reject-time-aware algorithms that can efficiently explore this expanded model-selection space is an interesting direction for future work.

\textbf{Dynamic Key Insertion and Deletion in CLBF.}
While CLBF can procedurally support key insertion by following \cref{alg:key_insertion}, updates may cause distribution shift and reduce the optimality of the chosen configuration, degrading accuracy.
Likewise, CLBF can support key deletion by replacing Bloom-filter components with deletion-capable AMQs (e.g., Counting Bloom filters or Cuckoo filters), but deletions can similarly make the configuration suboptimal and hurt accuracy.
A practical approach is to monitor performance and periodically rebuild CLBF once the update volume or measured degradation exceeds a predefined threshold.
Developing incremental maintenance strategies with formal guarantees and low rebuild overhead is an important direction for future work.

\section{Related Work}
\label{sec:related_work}

Bloom filter~\cite{bloom1970space} is one of the most fundamental and widely used data structures for approximate membership queries. 
Its fast query performance and low memory usage make it valuable in memory-constrained and latency-sensitive scenarios such as networks~\cite{broder2004network, tarkoma2011theory, geravand2013bloom}, databases~\cite{chang2008bigtable, goodrich2011invertible, lu2012bloomstore, kipf2020cukoo}, and information retrieval~\cite{jain2005using, li2025geobloom}.
A Bloom filter storing $n$ keys with FPR $f$ uses about $(\log_2 e) \cdot n\log_2(1/f)$ bits, which is $\log_{2}(e)$ times the theoretical lower bound of $n\log_{2}(1/f)$~\cite{pagh2005optimal}.
Several improved versions of the Bloom filter, such as the Cuckoo filter~\cite{fan2014cuckoo}, Vacuum filter~\cite{wang2019vacuum}, Xor filter~\cite{graf2020xor}, and Ribbon filter~\cite{peter2021ribbon}, have been proposed to get closer to the theoretical lower bound.

LBFs incorporate a learned model as part of the membership test to reduce the total memory usage~\cite{kraska2018case}. 
Following the naive LBF design, Sandwiched LBF~\cite{mitzenmacher2018model} improves memory efficiency by adding a prefilter before model inference and optimizing the allocation between the prefilter and the backup filter. 
Ada-BF~\cite{dai2020adaptive} and PLBF~\cite{vaidya2021partitioned} further exploit the model score by partitioning queries into score ranges and assigning different backup filters to different ranges, enabling finer-grained use of model confidence and improving the trade-off between memory usage and FPR.

Optimizing the hash functions is another approach to improving memory efficiency~\cite{kraska2018case}.
HABF~\cite{xie2021hashadaptive,li2023pareto} uses a lightweight data structure called HashExpressor to select suitable hash functions for each key, and PHBF~\cite{bhattacharya2022new} employs projections as hash functions.

While many studies optimize Bloom-filter configurations for a fixed trained model, some investigate the choice of the ML model itself~\cite{fumagalli2022choice, dai2022optimizing, malchiodi2024role}.
They evaluate different models and LBF configurations across datasets, suggesting that the best model depends on factors such as dataset noisiness and the importance of reject time.
However, these works provide only qualitative guidelines based on empirical trends and do not propose a method for automatically selecting the optimal model size.
Concurrently with our work, Shang et al.~\cite{shang2025gama} proposed Gama, a general framework for adaptively optimizing memory allocation between the ML model and filter components of an LBF.
As discussed in \cref{sec:discussion}, Gama formulates configuration selection as a generic black-box optimization problem and primarily targets the memory--FPR trade-off, whereas CLBF exploits its cascade structure and explicitly incorporates reject time into its optimization objective.

Cascaded designs that combine multiple filters in stages have also been proposed~\cite{salikhov2014using,engljahringer2022cascaded,deeds2020stacked}.
However, these approaches differ substantially from CLBF in both problem setting and architecture.
Traditional cascaded filters do not use machine learning models; they adopt cascading to remove false positives caused by known non-keys in stages.
CLBF, by contrast, uses an ML model to suppress false positives not only for known non-keys but also for non-keys that resemble them.
Accordingly, existing cascaded filters alternate layers that store keys and non-keys, whereas the internal Bloom filters in CLBF always store keys only.

Concurrently with our work, ELBF~\cite{lin2025ensemble} was proposed.
ELBF selects a subset of weak learners from a candidate pool and allocates space to the corresponding backup Bloom filters.
Whereas ELBF adopts a parallel design and primarily optimizes the trade-off between memory usage and FPR, CLBF adopts a serial cascade and jointly optimizes model size and filter allocation while explicitly optimizing reject time.
In our experiments, we implement ELBF using the authors' recommended hyperparameters and observe that CLBF achieves better memory--FPR and memory--reject-time trade-offs across datasets (and ML model variants; see \cref{sec:appendix:ml_model_variants}).

Overall, CLBF differs from prior LBF designs by jointly optimizing model size, Bloom-filter allocation, and reject time within a unified cascaded architecture.

\section{Conclusion}
\label{sec:conclusion}
We proposed CLBF to address two critical issues in existing LBFs.
(1) By training a large ML model and then reducing its size appropriately, CLBF achieves a near-optimal balance between model and filter size, minimizing overall memory usage.
(2) By branching on tentative scores and using intermediate Bloom filters, CLBF significantly reduces reject time.
CLBF broadens the applicability of LBFs and provides a foundation for further progress on these issues.


\clearpage
\bibliographystyle{ACM-Reference-Format}
\bibliography{main}

\clearpage
\appendix
\section{Explicit Definition of \texorpdfstring{$\mathcal{L}_d$}{L\_d}}
\label{sec:appendix:L_d_def}

Recall that throughout \cref{sec:dynamic_programming_for_optimizing_d_and_f}, the intermediate threshold vector $\bm{\theta}$ is fixed and therefore omitted from the notation.
This appendix gives the explicit definition of the suffix objective $\mathcal{L}_d(D,\mathcal{F})$ used in \cref{eq:dp}.

We define the suffix memory usage $M_d(D,\mathcal{F})$ and suffix reject time $R_d(D,\mathcal{F})$ by restricting the summations in \cref{eq:memory_size,eq:reject_time} to depths $i \ge d$:
\begin{align}
\label{eq:memory_size_d}
    M_d(D,\mathcal{F}) &\coloneq \sum_{i=d}^{D} \mathrm{Size}(\mathrm{ML}_i) + \sum_{i=d}^{D} \mathrm{Size}(\mathrm{TBF}_i) + \nonumber \\
    &\qquad + \sum_{i=d}^{D-1} \mathrm{Size}(\mathrm{BBF}_i) + \sum_{k=1}^{K} \mathrm{Size}(\mathrm{FBF}_k), \\
\label{eq:reject_time_d}
    R_d(D,\mathcal{F}) &\coloneq \sum_{i=d}^{D} \left(\mathrm{Time}(\mathrm{ML}_i) \cdot h^{(\mathrm{t})}_i \prod_{j=1}^{i} f^{(\mathrm{t})}_j\right).
\end{align}
The Bloom-filter size terms $\mathrm{Size}(\mathrm{TBF}_i)$, $\mathrm{Size}(\mathrm{BBF}_i)$, and $\mathrm{Size}(\mathrm{FBF}_k)$ are as in \cref{eq:bloom_filter_size_1,eq:bloom_filter_size_2,eq:bloom_filter_size_3}.
The suffix objective is then defined analogously to \cref{eq:objective}:
\begin{equation}
    \label{eq:L_d}
    \mathcal{L}_d(D,\mathcal{F}) \coloneq \lambda \cdot \frac{M_d(D,\mathcal{F})}{M_\mathrm{BF}} + (1-\lambda) \cdot \frac{R_d(D,\mathcal{F})}{R_\mathrm{BF}}.
\end{equation}
By definition, $\mathcal{L}_1(D,\mathcal{F}) = \mathcal{L}(D,\mathcal{F})$, and $\mathrm{dp}(1,1)$ corresponds to the original optimization problem.

\section{Effect of ML Model Variants}
\label{sec:appendix:ml_model_variants}

\begin{figure*}
    \centering
    \includegraphics[width=\textwidth]{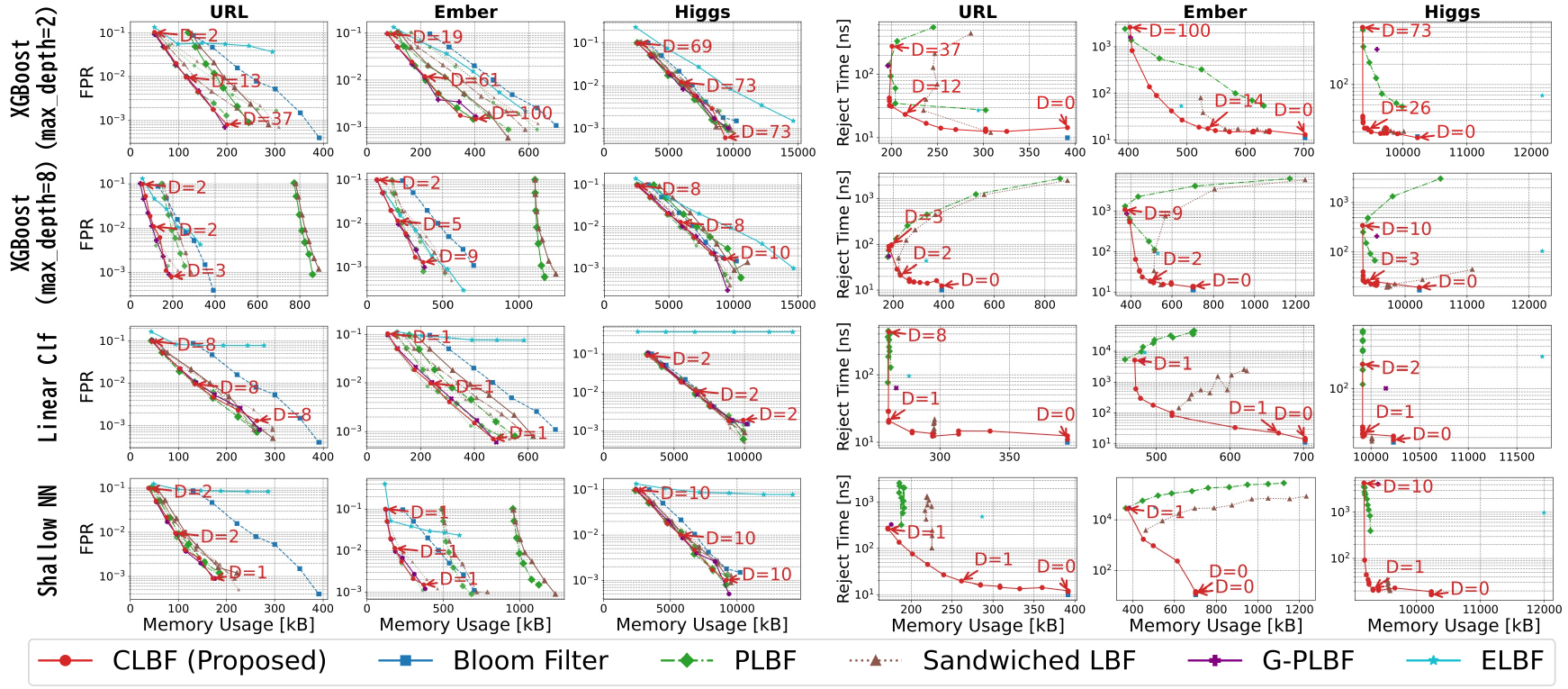}
    \caption{Effect of different ML models on CLBF: Across various architectures and hyperparameter settings, CLBF consistently achieves superior memory--FPR and memory--reject-time trade-offs compared to existing LBFs.}
    \label{fig:appendix:ml_model_variants}
    \Description{Across different ML models, CLBF consistently achieves better trade-offs between memory usage, accuracy, and reject time than existing LBFs.}
\end{figure*}

Complementing the main experiments in \cref{sec:experiments}, which use XGBoost with \texttt{max\_depth}$=4$ and $\bar{D}=100$, this section evaluates whether CLBF's advantages persist under alternative ML architectures and XGBoost hyperparameter settings.
Throughout this section, we fix the maximum number of weak learners for CLBF to $\bar{D} = 10$; for the other LBFs, we report results for $D \in \{1, 5, 10\}$ in the memory--FPR plots and for $D \in \{1, 2, \dots, 10\}$ in the memory--reject-time plots.
Since this experiment focuses on varying the ML model, we exclude PHBF and HABF, which do not employ an ML model.
As a hyperparameter ablation for XGBoost, we vary \texttt{max\_depth} $\in \{2, 4, 8\}$: a smaller \texttt{max\_depth} yields smaller weak learners and faster inference, but reduces discriminative power.
As an ablation on the model architecture itself, we additionally consider a linear classifier (\textbf{Linear Clf}) and a shallow neural network (\textbf{Shallow NN}), both trained in a boosting framework:
\begin{itemize}[leftmargin=1.5em]
    \item \textbf{Linear classifier}: We trained linear models with a learning rate of $10^{-3}$ for $50$ epochs.
    \item \textbf{Shallow neural network}: We trained one-hidden-layer neural networks with hidden dimension $10$ and a learning rate of $10^{-3}$ for $50$ epochs.
\end{itemize}

The left part of \cref{fig:appendix:ml_model_variants} shows the trade-off between memory usage and FPR under each hyperparameter/architecture setting.
Across all cases, CLBF achieves a better trade-off than existing LBFs.
The optimal number of weak learners depends on the hyperparameters and architectures of the underlying ML model:
for less expressive models such as XGBoost ($\texttt{max\_depth}=2$) or a linear classifier, a single learner ($D=1$) is often insufficient and using more learners improves memory efficiency,
whereas for more expressive models such as XGBoost ($\texttt{max\_depth}=8$) or a shallow NN, a small number of learners tends to be optimal.
In all settings, CLBF automatically selects an appropriate number of weak learners and consistently achieves better memory efficiency than existing LBFs.

The right part of \cref{fig:appendix:ml_model_variants} shows the trade-off between memory usage and reject time under each hyperparameter/architecture setting.
Again, CLBF consistently achieves a superior trade-off between memory usage and reject time compared to existing LBFs.
In particular, on the Higgs dataset, the Shallow NN yields the smallest memory usage, and the gap in reject time remains large: at a memory usage of $9{,}190\,\mathrm{kB}$, the reject time of CLBF is about 65$\times$ shorter than that of PLBF.

\section{Comprehensive Experiments on All Datasets}
\label{sec:appendix:comprehensive}

Complementing the detailed discussion in \cref{sec:experiments} on three representative benchmarks (URL, Ember, and Higgs), here we report the same experimental protocol across all 14 datasets listed in \cref{tab:appendix:datasets}, including nine Kitsune network-intrusion splits~\cite{mirsky2018kitsune}, MNIST~\cite{deng2012mnist}, and DNA~\cite{rahman2021representation}.
The compared methods, metrics, and hyperparameters are identical to those in \cref{sec:experiments}.

Additional datasets are as follows:
\begin{itemize}[leftmargin=1.5em]
    \item \textbf{Kitsune [Active, Fuzzing, Mirai, ARP, OS, SSL, Video, SSDP, SYN]}~\cite{mirsky2018kitsune}: nine network intrusion detection splits with $115$-dimensional traffic features.
    We treat label $1$ (attack) as keys and label $0$ (benign) as non-keys.
    \item \textbf{MNIST}~\cite{deng2012mnist}: $28{\times}28$ grayscale digit images.
    Three digit classes are chosen at random as keys and the remaining classes as non-keys (784-dimensional inputs after flattening).
    \item \textbf{DNA}~\cite{rahman2021representation}: $14$-mers from human chromosome~14.
    We use $49{,}906{,}253$ distinct $14$-mers as keys and sample the same number of non-keys uniformly from the remaining strings over $\{\mathtt{A}, \mathtt{T}, \mathtt{C}, \mathtt{G}\}$.
    Each $14$-mer is encoded as a $14$-dimensional vector by mapping $\mathtt{A}, \mathtt{T}, \mathtt{C}, \mathtt{G}$ to integers $0$--$3$.
\end{itemize}

\begin{table}[t]
    \centering
    \caption{Statistics of the datasets. For all datasets, all elements in $\mathcal{S}$ are used for training and calibration, and $90\%$ of the elements in $\mathcal{Q}$ are used for training while the remaining $10\%$ are used for calibration. We reserve $10{,}000$ non-keys that are not used in either training or calibration as a test set.}
    \label{tab:appendix:datasets}
    \setlength{\tabcolsep}{4pt}
    \small
    \begin{tabular}{@{}lrrrr@{}}
        \toprule
        Dataset & dim. & $|\mathcal{S}|$ & $|\mathcal{Q}|$ & \#non-key (test)\\
        \midrule
        URL & 20 & 223{,}088 & 418{,}103 & 10{,}000 \\
        Ember & 2{,}381 & 400{,}000 & 390{,}000 & 10{,}000 \\
        Higgs & 28 & 5{,}829{,}123 & 5{,}160{,}877 & 10{,}000 \\
        Kitsune [Active] & 115 & 923{,}216 & 1{,}345{,}475 & 10{,}000 \\
        Kitsune [Fuzzing] & 115 & 432{,}783 & 1{,}801{,}358 & 10{,}000 \\
        Kitsune [Mirai] & 115 & 642{,}516 & 111{,}621 & 10{,}000 \\
        Kitsune [ARP] & 115 & 1{,}145{,}272 & 1{,}348{,}997 & 10{,}000 \\
        Kitsune [OS] & 115 & 65{,}700 & 1{,}622{,}153 & 10{,}000 \\
        Kitsune [SSL] & 115 & 92{,}652 & 2{,}104{,}921 & 10{,}000 \\
        Kitsune [Video] & 115 & 102{,}499 & 2{,}359{,}904 & 10{,}000 \\
        Kitsune [SSDP] & 115 & 1{,}439{,}604 & 2{,}627{,}664 & 10{,}000 \\
        Kitsune [SYN] & 115 & 7{,}038 & 2{,}754{,}240 & 10{,}000 \\
        MNIST & 784 & 21{,}604 & 38{,}396 & 10{,}000 \\
        DNA & 14 & 49{,}906{,}253 & 49{,}906{,}253 & 10{,}000 \\
        \bottomrule
    \end{tabular}
\end{table}

\begin{figure*}[t]
    \centering
    \includegraphics[width=\textwidth]{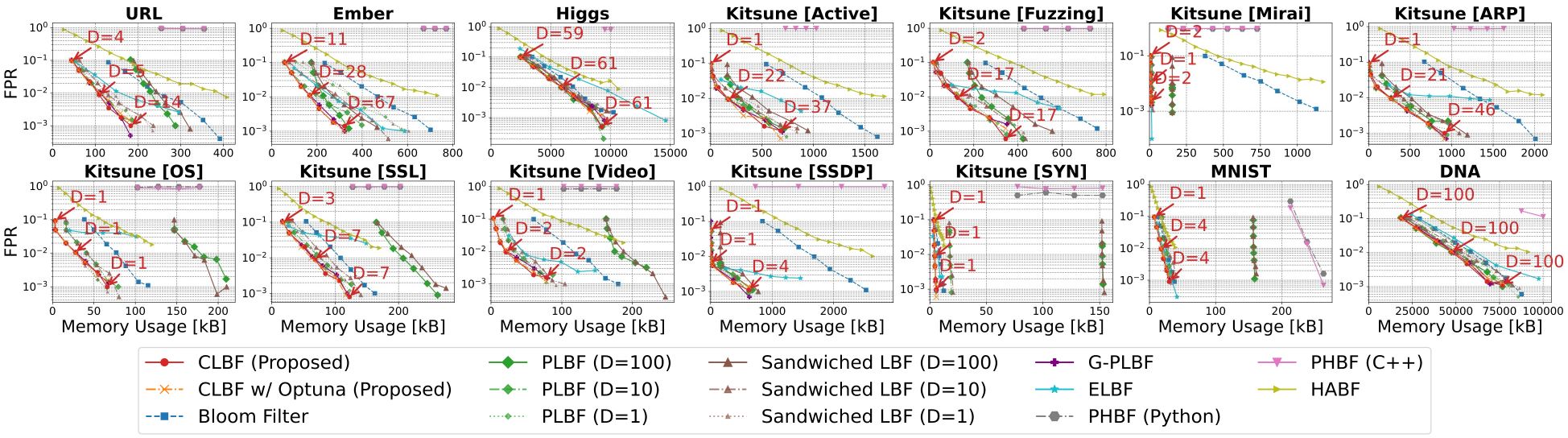}
    \caption{Memory usage vs.\ FPR across all datasets (XGBoost): CLBF matches or improves upon the best prior LBF configurations at the same target FPR.}
    \label{fig:appendix:memory_fpr}
    \Description{Multi-panel plot of memory versus false positive rate for many datasets, showing CLBF near the Pareto frontier of learned Bloom filters.}
\end{figure*}

\begin{figure*}[t]
    \centering
    \includegraphics[width=\textwidth]{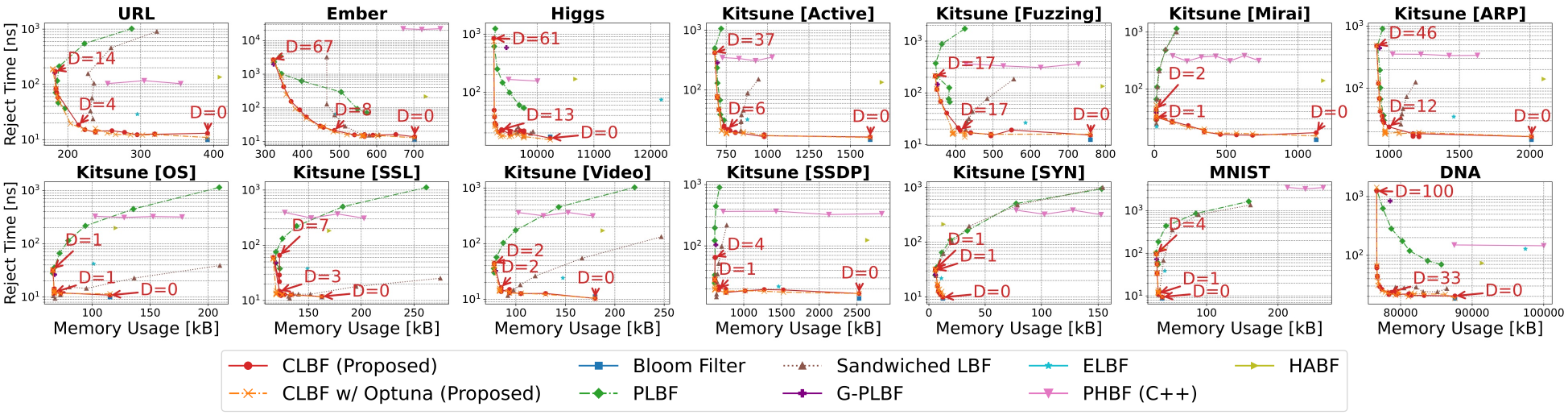}
    \caption{Memory usage vs.\ reject time for non-keys across all datasets (XGBoost): CLBF reduces reject time relative to PLBF by using cascaded early rejection, while maintaining strong memory efficiency.}
    \label{fig:appendix:memory_query_time}
    \Description{Multi-panel plot of memory versus average reject time for non-key queries across datasets, showing CLBF with lower reject times than PLBF at similar memory.}
\end{figure*}

\begin{figure*}[t]
    \centering
    \includegraphics[width=\textwidth]{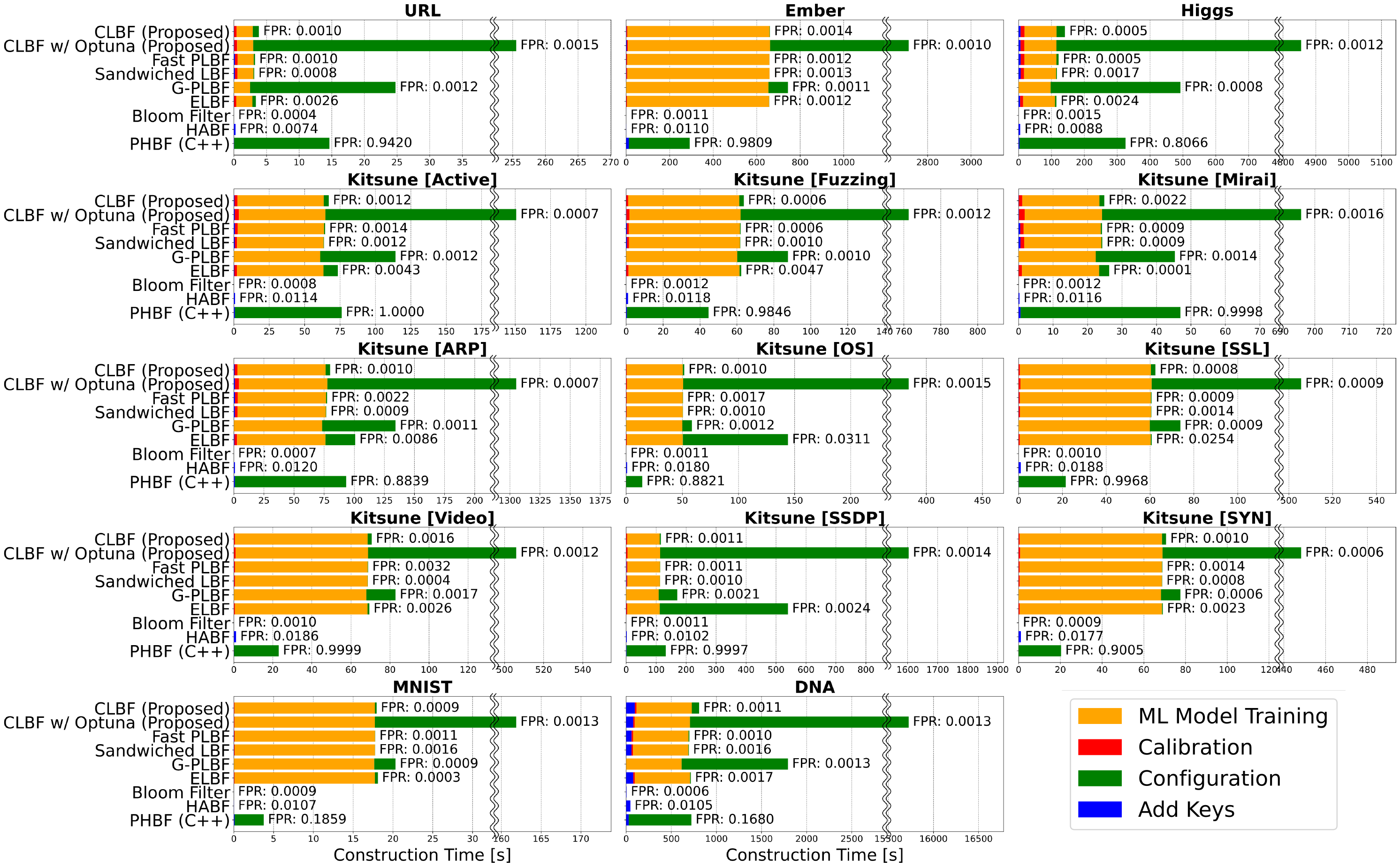}
    \caption{Construction time across all datasets: CLBF remains comparable to existing LBFs (at most about 1.23$\times$ overhead) despite searching for a near-optimal configuration.}
    \label{fig:appendix:time_hist_all}
    \Description{Multi-panel comparison of construction times for multiple learned Bloom filters across many datasets, showing CLBF with construction time comparable to existing LBFs.}
\end{figure*}

\begin{figure*}[t]
    \centering
    \includegraphics[width=\textwidth]{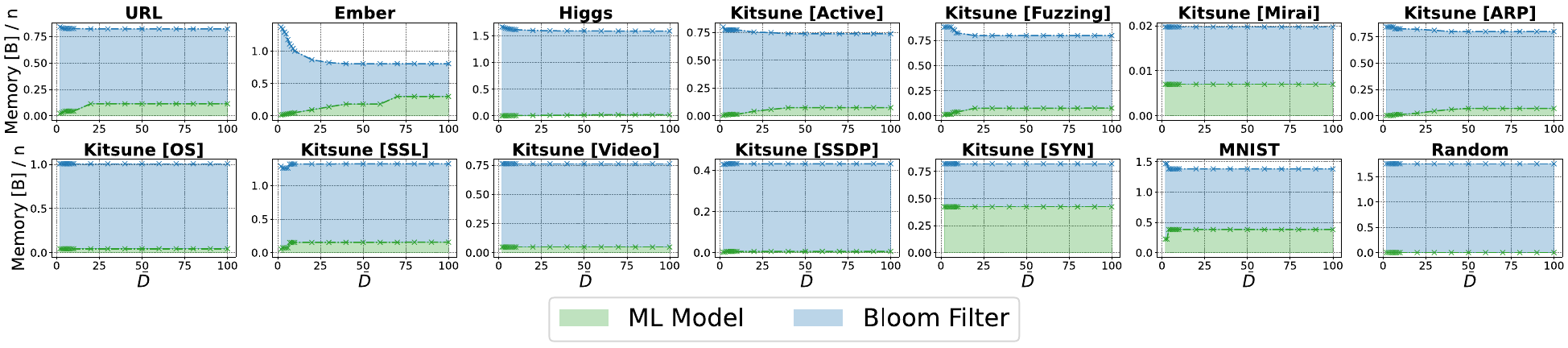}
    \caption{Effect of the number of given models $\bar{D}$ in CLBF across all datasets ($F=0.001$): total memory decreases then stabilizes as CLBF selects an optimal subset of weak learners.}
    \label{fig:appendix:D_clbf}
    \Description{Multi-panel plots of memory versus number of available weak learners for CLBF, showing decreasing then flat curves with a marked chosen subset size.}
\end{figure*}

\begin{figure*}[t]
    \centering
    \includegraphics[width=\textwidth]{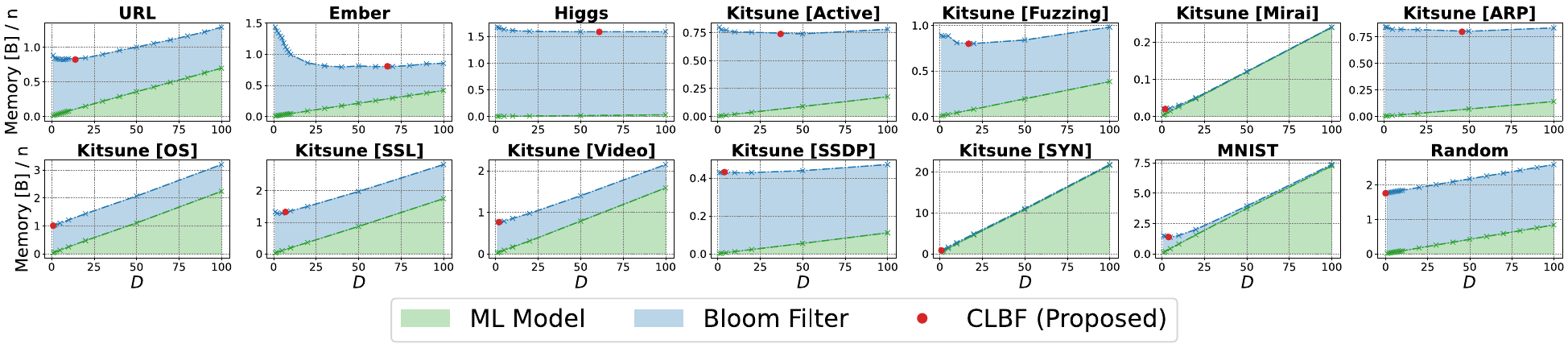}
    \caption{Effect of the number of models $D$ in PLBF across all datasets ($F=0.001$): memory initially decreases but eventually increases because PLBF must keep every weak learner.}
    \label{fig:appendix:D_plbf}
    \Description{Multi-panel plots of memory versus number of weak learners for PLBF, showing U-shaped total memory as models accumulate.}
\end{figure*}

\subsection{Memory--FPR and Memory--Reject-Time Trade-offs}

\cref{fig:appendix:memory_fpr,fig:appendix:memory_query_time} show the memory--FPR and memory--reject-time trade-offs over every dataset.
Qualitatively, the trends summarized in \cref{sec:experiments} hold broadly: CLBF's operating points tend to lie on or below the Pareto frontier of existing LBFs in the memory--FPR plane, and CLBF achieves substantially shorter reject times than PLBF at comparable memory budgets, with Sandwiched LBF often exhibiting competitive reject times but worse memory efficiency.
The additional Kitsune splits illustrate that these behaviors are stable under varying key/non-key imbalance and traffic regimes (e.g., small key sets such as Kitsune [SYN] vs.\ large non-key sets such as Kitsune [Fuzzing]).
On very large key sets (Higgs, DNA), CLBF still traces favorable trade-off curves, indicating that the dynamic-programming construction scales to demanding cardinality regimes without losing its advantage over fixed-structure LBFs.
The gains can be dramatic when the classification problem is easy: on Kitsune [SSDP] with $F = 0.01$, CLBF uses only 3.3\,kB of memory, whereas a standard Bloom filter requires 1{,}684\,kB, corresponding to a 99.8\% reduction.
We also observe that Bayesian optimization of the intermediate thresholds can be more effective on some of these datasets than on the three representative ones; on Kitsune [ARP], it reduces the reject time by up to 35\%.

\subsection{Construction Time}

\cref{fig:appendix:time_hist_all} compares construction time for representative LBF configurations across all datasets.
Consistent with the three-dataset discussion for URL, Ember, and Higgs in the main text, CLBF incurs only a mild additional overhead relative to Sandwiched LBF---about 1.003$\times$ to 1.23$\times$---despite searching for a near-optimal configuration.
The construction time of G-PLBF is substantially longer than that of CLBF because G-PLBF relies on generic black-box optimization with sequential evaluation of candidate configurations.
The trade-off offered by CLBF w/ Optuna is also visible here.
For example, on Kitsune [ARP], CLBF w/ Optuna reduces the reject time from 30\,ns to 20\,ns compared to CLBF, but the construction time increases from 90 to 1{,}800 seconds, which illustrates why we recommend CLBF as the practical default.

\subsection{Ablation on the Number of ML Models}

\cref{fig:appendix:D_clbf,fig:appendix:D_plbf} extend the ablation in \cref{sec:exp:ablation_study_on_D} to all datasets.
For CLBF, increasing the number of available weak learners $\bar{D}$ reduces total memory until sufficient capacity is reached; beyond that point, optimization selects an optimal number of weak learners $D \le \bar{D}$ and the total size plateaus because unused models are pruned.
For PLBF, larger $D$ necessarily retains all weak learners, so memory eventually rises as models accumulate even when additional trees no longer improve the backup Bloom filter as much.
Across datasets, including high-dimensional (Ember, MNIST) and massively scaled key sets (DNA), CLBF's selected $D$ (highlighted in the plots) tracks the region where total memory is near-minimal, supporting the claim that the cascaded construction automatically balances model size and filter size.

\end{document}